 \documentclass[a4paper,12pt]{article}

\usepackage[a4paper]{geometry}

  \usepackage{natbib}
  \usepackage[figuresright]{rotating}
\usepackage{mathptmx} 
\usepackage{amsmath}
\usepackage{amsfonts}
\usepackage{amssymb} 
  \usepackage{amsthm}
\usepackage{url}
  \usepackage{graphicx}
\graphicspath{{/}{images/}}

\usepackage{color}

\definecolor{lightred}{rgb}{0.5,0,0}
\definecolor{lightblue}{rgb}{0,0,0.5}
\definecolor{mgrey}{rgb}{0.5,0.5,0.5}

\long\def\@makefntext#1{\raggedright
\@setpar{\@@par\@tempdima \hsize
 \advance\@tempdima-\@footindent
 \parshape \@ne \@footindent \@tempdima}\par
 \noindent \hbox to \z@{\hss\@thefnmark\enskip}#1}

 \usepackage{makeidx}
 \makeindex

  \theoremstyle{definition}

  \title{Market Microstructure Knowledge Needed for Controlling an Intra-Day Trading Process}

  \author{Charles-Albert Lehalle\thanks{Global Head of Quantitative Research, Cr\' edit Agricole Cheuvreux, {\tt charles@lehalle.net}}}

\begin{document}
  \maketitle 

\def\rk#1{\emph{#1}}
\def\Esp{\mathbb{E}}
\def\Var{\mathbb{V}}
\def\one{{1\!\!1}}

\def\calT{{\cal T}}

\def\eps{\varepsilon}
\def\tC{{\tilde C}}

\def\bV{\bar{V}}
\def\bR{\mathbf{r}}

\def\calC{{\cal C}}
\def\calS{{\cal S}}

\def\sumN{\sum_{n=1}^N}
\def\sumL{\sum_{\ell=1}^N}
\def\bv{\mathbf{v}}

\def\X(#1){(x_{#1}-x_{#1+1})}

\def\df#1#2{\frac{\partial{#1}}{\partial{#2}}}

\def\rred#1{#1}


\begin{abstract} 
A \rred{great deal} of academic and theoretical work  \rred{has} been dedicated to 
optimal liquidation of large orders 
these last twenty years.
  The optimal split of an order through time (`optimal trade scheduling') and space (`smart order routing') 
is of high interest \rred{to} practitioners because of the increasing complexity 
of the market micro structure \rred{because of} 
\rred{the evolution recently} of regulations and liquidity worldwide. 
This chapter \rred{translates into} quantitative terms these
 regulatory issues and, more broadly, current market design.

  It \rred{relates} the recent advances in optimal trading, order-book simulation and optimal \rred{liquidity to} the 
reality of trading in an emerging global network of liquidity.
\end{abstract}

\tableofcontents
\clearpage

\section{Market Microstructure Modeling and Payoff Understanding are Key Elements of 
Quantitative Trading}\label{lehalle_sec1}

As  \rred{is well} known, optimal (or quantitative) trading is about finding the proper balance between 
providing liquidity \rred{in order} to minimize the impact of the trades, 
and consuming liquidity \rred{in order} to minimize the market risk exposure, 
while taking profit \rred{through  potentially} instantaneous trading signals, 
supposed to be triggered by liquidity inefficiencies.

The mathematical framework required to solve this kind of optimization \rred{problem} needs:
\begin{itemize}
\item a model of the 
consequences of the different ways \rred{of interacting} with liquidity (\rred{such as the} market impact model 
\citep{citeulike:4325901, Bouchaud06, citeulike:5177397});
\item  a proxy for the `market risk' (the most natural of 
them being the high frequency volatility \citep{AITJAC07,scales05,citeulike:8317402});
\item  and a model 
\rred{for quantifying} the likelihood of the liquidity state of the market 
\citep{citeulike:7344893,citeulike:8318790}. 
\end{itemize}
A utility function \rred{then allows these different effects to be
consolidated} with respect to the goal of the trader:
\begin{itemize}
\item  minimizing the impact of large trades under price, 
duration and volume constraints (typical for brokerage trading \citep{OPTEXECAC00}); 
\item providing as \rred{much} liquidity as possible under inventory constraints (typical for market-makers 
\cite{avst08} or \rred{\cite{GLFT}}); 
\item or following a belief \rred{about} the trajectory of the market 
(typical of arbitrageurs \citep{citeulike:5094012}).
\end{itemize}

Once these key elements \rred{have been} defined, rigorous mathematical optimization methods can be used to derive \rred{the} 
optimal behavior \citep{citeulike:5797837,citeulike:8531791}. Since the optimality of the result 
\rred{is strongly dependent} on the \rred{phenomenon being modeled}, \rred{some} 
understanding of the market microstructure is a 
prerequisite \rred{for ensuring} the applicability of a given theoretical framework.

\begin{figure}[!h]
\def\bido#1#2{%
\put(#1,0){\line(0,1){#2}}
\put(#1,#2){\line(1,0){.01}}
\put(#1,#2){\line(-1,0){.01}}
}
\def\bida#1#2#3{%
\put(#1,#2){\line(0,1){#3}}
\put(#1,#2){\line(1,0){.01}}
\put(#1,#2){\line(-1,0){.01}}
}
\def\asko#1#2{%
\put(#1,0){\line(0,-1){#2}}
\put(#1,-#2){\line(1,0){.01}}
\put(#1,-#2){\line(-1,0){.01}}
}

\setlength{\unitlength}{\textwidth}
\begin{picture}(1,0.6)(0,-0.3)
\thicklines
\put(0,0){\vector(0,1){.3}}
\put(0,0){\line(0,-1){.3}}
\put(0,.3){ Quantity}
\put(0,0){\vector(1,0){1}}
\put(0.95,0.025){Price}

\linethickness{1mm}

\color{blue}

\bido{.1}{.05}
\bido{.1}{.15}\put(0.0,0.10){ Buy}
\bido{.1}{.25}

\bido{.2}{.15}
\bido{.2}{.20}
\bido{.2}{.30}

\bido{.3}{.02}
\bido{.3}{.17}
\bido{.3}{.26}

\bido{.4}{.12}
\bido{.4}{.21}
\put(.35,.22){Best Bid}

\bida{.6}{.15}{.05}
\put(.6,.2){\line(1,0){.01}}
\put(.6,.2){\line(-1,0){.01}}
\put(.62,.22){A buy order at this price}
\put(.64,.19){generates a trade}

\color{red}

\asko{.6}{.07}\put(0.0,-0.10){ Sell}
\put(.55,-.10){Best Ask}

\asko{.7}{.05}
\asko{.7}{.16}
\asko{.7}{.21}

\asko{.8}{.09}
\asko{.8}{.19}
\asko{.8}{.27}

\asko{.9}{.10}
\asko{.9}{.18}
\asko{.9}{.28}

\color{black}

\thicklines
\put(.6,.13){\vector(0,-1){.12}}

\thinlines

\multiput(.4,0)(0,-.03){7}{\line(0,-1){.02}}
\multiput(.3,0)(0,-.03){7}{\line(0,-1){.02}}

\thicklines
\put(.5,-.05){\vector(1,0){.1}}
\put(.5,-.05){\vector(-1,0){.1}}
\put(.4,-.04){ bid-ask spread}

\put(.35,-.2){\vector(1,0){.05}}
\put(.35,-.2){\vector(-1,0){.05}}
\put(.3,-.23){tick size}

\end{picture}

\caption{\rred{Idealized} order-book}
  \label{fig:LOB}
\end{figure}
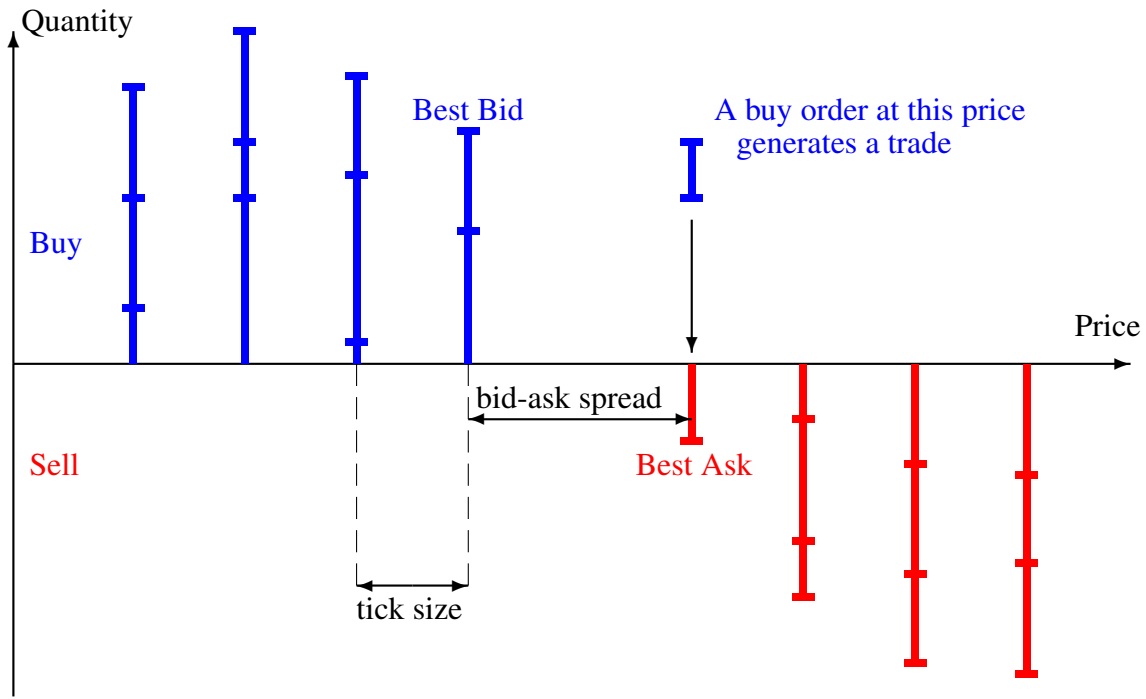

The \emph{market microstructure} is the ecosystem \rred{in which buying and selling interests meet}, 
giving birth to trades. Seen from outside the microstructure, the prices of the traded shares are often 
uniformly sampled to build time series that are modeled via martingales \citep{citeulike:1681881} 
or studied using econometrics. Seen from the inside of electronic markets, buy and sell open interests 
(i.e. passive limit orders) form \emph{limit order books}, where an impatient trader can find two 
different prices: the highest of the resting buy orders if he needs to sell, 
and the lowest of the selling ones if he needs to buy (\rred{see} Figure \ref{fig:LOB}). 
The \rred{buying and selling} price are thus not equal. Moreover, the price will monotonically increase 
(for impatient buy orders) or decrease (for impatient sell orders) with  the quantity to trade, following a 
concave function \citep{farmer03a}: the more you trade, the \rred{worse the price you will get}.

The market microstructure is \rred{strongly} conditioned by the \emph{market design}:
\begin{itemize}
\item the set of 
explicit rules governing the \emph{price formation process} (PFP);
\item the type of auction (fixing or continuous ones); 
the tick size (i.e. the minimum allowed difference between two consecutive prices); 
\item the interactions between trading platforms (such as `trade-through rules', pegged orders, interactions between 
visible and hidden orders, etc.);
\end{itemize}
are typical elements of the market design.

The market microstructure of an asset class is a mix of the market design, the trading behaviors of trading agents, 
the regulatory environment, and the availability of correlated instruments (such as Equity Traded Funds, Futures or 
any kind of derivative products). 
Formally, the microstructure of a market can be seen as several sequences of auction mechanisms taking place
 in parallel, each of them having its \rred{own particular characteristics}. For instance the German market place 
is mainly composed (as of 2011) of the Deutsche B\"orse regulated market, the Xetra mid-point, the Chi-X visible order book, 
Chi-delta (the Chi-X hidden mid-point), Turquoise Lit and Dark Pools, BATS pools. 
The regulated market implements a sequence of fixing auctions and continuous auctions (one open fixing, 
one continuous session, one mid-auction and one closing auction); others implement only continuous auctions, 
and Turquoise mid-point implements optional random fixing auctions.

To optimize his behavior, a trader has to choose an abstract description of the microstructure of the 
markets he will interact with: this will be his model of market microstructure. It can be a statistical
 `macroscopic' one as in the widely-used Almgren--Chriss framework \citep{OPTEXECAC00}, in which the time 
is sliced \rred{into intervals of 5 or 10 minutes  duration} during which the interactions with the market 
\rred{combine} two statistical phenomena: 
\begin{itemize}
\item the market impact as a function of the `participation rate' of 
the trader; 
\item and the volatility as a proxy of the market risk. 
\end{itemize}
It can also be a microscopic description 
of the order book behavior as in the Alfonsi--Schied proposal \citep{citeulike:6615020} in which the shape 
of the order book and its resilience to liquidity-consuming orders is modeled.

This chapter will thus \rred{describe} some relationships between the market design and the market microstructure 
using \rred{European and American examples} since they have seen regulatory changes (in 2007 for Europe with the MiFI 
Directive, and in 2005 for the USA with the \rred{NMS regulation}) 
as much as behavioral changes (with the financial crisis of 2008).
A detailed description of some important elements of the market microstructure will be conducted: 
\begin{itemize}
\item dark pools;
\item impact of fragmentation on the price formation process;
\item tick size;
\item auctions, etc. 
\end{itemize}
Key events like the 
6 May 2010 flash crash in the US market and some European market outages will \rred{also receive attention}.

To obtain an optimal trading trajectory, a trader needs to define its payoff. Here also, choices have to be made 
from a mean-variance \rred{criterion} \citep{OPTEXECAC00} to stochastic impulse control \citep{citeulike:5797837} 
going through stochastic algorithms \citep{citeulike:5177512}.
This chapter \rred{describes the} statistical viewpoint of the Almgren--Chriss framework, showing how practitioners 
can use it to take into account a large variety of effects. It ends with comments on an order-flow oriented 
view of optimal execution, dedicated to smaller time-scale problems, \rred{such as} `\emph{Smart Order Routing}' (SOR).

\section{From Market Design to Market Microstructure: Practical Examples}\label{lehalle_sec2}

The recent history of the French equity market is archetypal in the sense that it went from a highly 
concentrated design with only one electronic platform \rred{hosted} in Paris \citep{MUN03} 
to a fragmented pan-European one with four visible trading pools and more than twelve `dark ones', 
located in London, in less than four years.

\begin{figure}[!h]
  \centering
  \includegraphics[width=\textwidth]{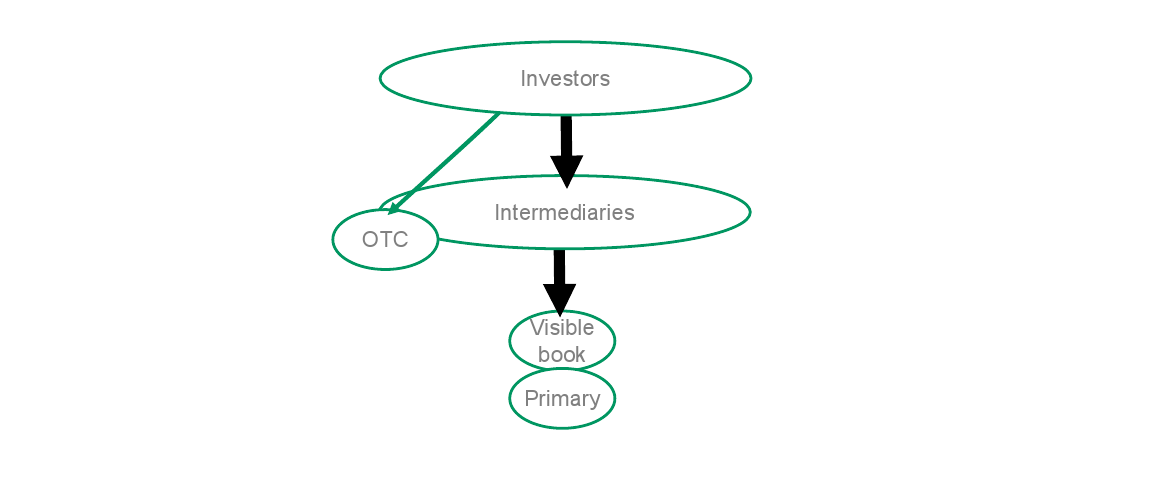}
  \caption{\rred{Idealized} pre-fragmentation market microstructures}
  \label{fig:MMS}
\end{figure}

Seen by economists and from \rred{outside the microstructure}, the equity market is a place 
where \emph{listed firms} raise capital offering shares \rred{for sale}. Once shares are available \rred{for 
buying and selling} in the market place, the mechanism of balance between offer and demand (in terms of 
intentions to buy and intentions to sell) forms a \emph{fair price}.

At the microstructure scale, the market place is more sophisticated. Market participants are no 
\rred{longer} just listed firms and investors making rational investment decisions; 
microstructure focuses on the process \rred{that allows investors to buy from, or sell to, one another}, 
putting emphasis on the \emph{Price Formation Process}, also \rred{known as} \emph{Price Discovery}. 
Moreover, recent regulations promote the use of electronic markets, \rred{since they are compatible} with the 
recording and traceability levels \rred{such markets provide}, leading to \emph{fragmented markets}. 
It is worthwhile  \rred{differentiating} between two states of the microstructure: 
\rred{pre- and post-fragmentation}, see Figure~\ref{fig:MMS} and \ref{fig:MMSfrag}: 
\begin{itemize}
\item \emph{Pre-fragmented microstructure}: before Reg NMS in the US and MiFID in Europe, 
the microstructure \rred{can} be \rred{pictured as} three distinct layers:
  \begin{itemize}
  \item investors, taking buy or sell decisions;
  \item intermediaries, giving \rred{unbiased advice} (through financial analysts or strategists) 
and providing access to trading pools they are members of; low frequency market makers (or maker-dealers) 
can be considered to be part of this layer;
  \item market operators: hosting the trading platforms, NYSE Euronext, NASDAQ, BATS, Chi-X, belong to this layer. 
They are providing matching engines to other market participants, hosting the \emph{Price Formation Process}.
  \end{itemize}

These three layers are simply connected: intermediaries concentrate a fraction of the buying and selling flows 
in a (small) \emph{Over the Counter} (OTC) market, the remaining open interests are \rred{placed}
 in the order books of 
the market operators. Facilitators (i.e. low frequency market makers or specialists), localized in the same layer
 \rred{as} the intermediaries, provide liquidity, thus minimizing the \emph{Market Impact} of \rred{orders from 
under-coordinated} investors (i.e. when a large buyer comes \rred{to} the market two hours after a large seller, any 
liquidity provider \rred{that is} able to sell to the first one and buy to the later will prevent a price oscillation; 
on the one hand he will be `rewarded' for this service through the bid--ask spread he will demand \rred{of} the two investors; 
on the other hand he will take the risk of a large change \rred{in} the \emph{fair price} \rred{that is} 
 in between the two transactions
 \citep{citeulike:7360166}, see Figure \ref{fig:syncorders}).

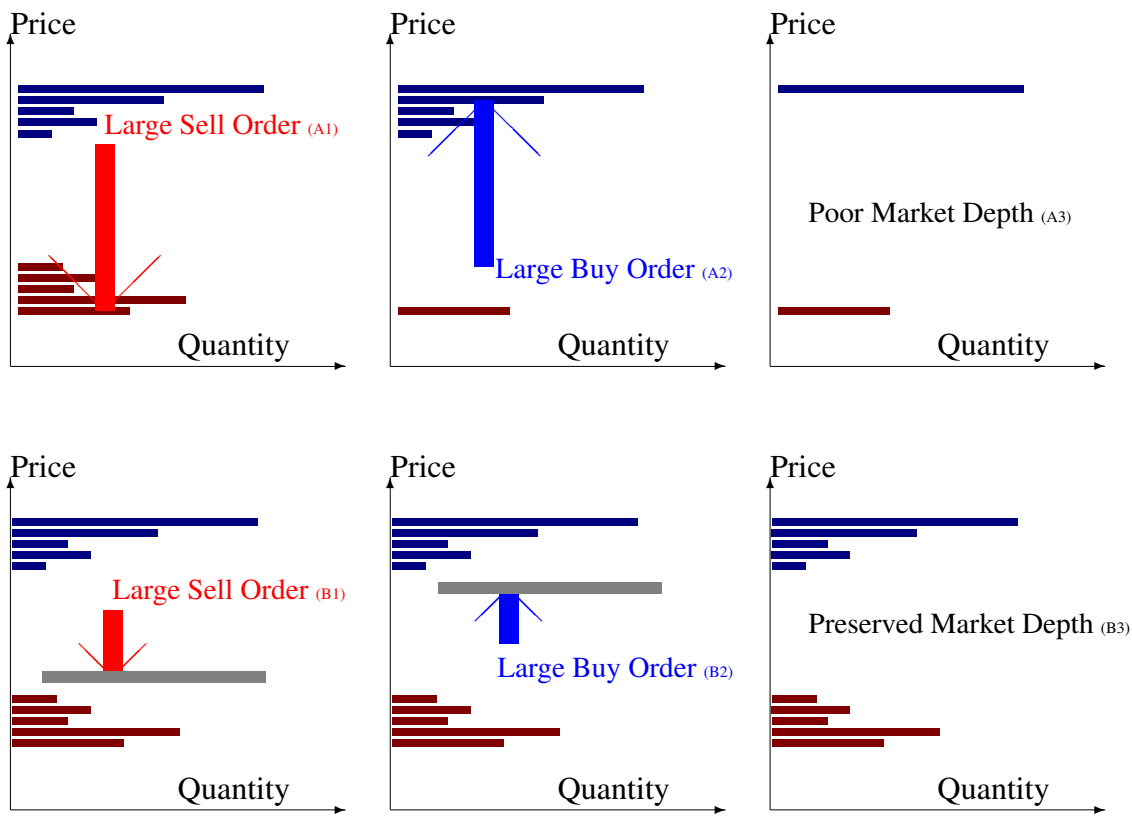
\begin{figure}[!h]



\def\steppob{.34}
\def\steppobtw{.68}
\setlength{\unitlength}{\textwidth}
\begin{picture}(1,0.7)(0,-0.4)
  \linethickness{0.1mm}
  \multiput(0,0)(\steppob,0){3}{
    \put(0,0){\vector(1,0){.3}}\put(.15,0.01){Quantity}
    \put(0,0){\vector(0,1){.3}}\put(0,.3){Price}

    \put(0,-0.4){\vector(1,0){.3}}\put(.15,-0.39){Quantity}
    \put(0,-0.4){\vector(0,1){.3}}\put(0,-0.1){Price}
  }
  \linethickness{1mm}
  \color{lightblue}
  \multiput(0,0)(\steppob,0){2}{
    \put(0,0.21){\line(1,0){.03}}
    \put(0,0.22){\line(1,0){.07}}
    \put(0,0.23){\line(1,0){.05}}
    \put(0,0.24){\line(1,0){.13}}
    \put(0,0.25){\line(1,0){.22}}
  }
  \put(\steppobtw,0.25){\line(1,0){.22}}
  
  \color{lightred}
  \put(0,0.05){\line(1,0){.10}}
  \put(0,0.06){\line(1,0){.15}}
  \put(0,0.07){\line(1,0){.05}}  
  \put(0,0.08){\line(1,0){.07}}
  \put(0,0.09){\line(1,0){.04}}

  \put(\steppob,0.05){\line(1,0){.10}}
  \put(\steppobtw,0.05){\line(1,0){.10}}

  \linethickness{2.5mm}
  \color{red}
  \put(0.07,0.20){\line(0,-1){.15}}
  \put(0.07,0.05){\line(-1,1){.05}}\put(0.07,0.05){\line(1,1){.05}}
  \put(0.07,0.21){\small Large Sell Order {\tiny (A1)}}
  \color{blue}
  \put(0.41,0.24){\line(0,-1){.15}}
  \put(0.41,0.24){\line(-1,-1){.05}}\put(0.41,0.24){\line(1,-1){.05}}
  \put(0.42,0.08){\small Large Buy Order {\tiny (A2)}}
  \color{black}
  \put(0.7,0.13){\small Poor Market Depth {\tiny (A3)}}
  \put(0.7,-0.24){\small Preserved Market Depth {\tiny (B3)}}

  \linethickness{1mm}
  \multiput(-0.02,0.05)(\steppob,0){3}{
    \color{lightblue}
    \put(0,-0.23){\line(1,0){.03}}
    \put(0,-0.22){\line(1,0){.07}}
    \put(0,-0.21){\line(1,0){.05}}
    \put(0,-0.20){\line(1,0){.13}}
    \put(0,-0.19){\line(1,0){.22}}

    \color{lightred}
    \put(0,-0.39){\line(1,0){.10}}
    \put(0,-0.38){\line(1,0){.15}}
    \put(0,-0.37){\line(1,0){.05}}  
    \put(0,-0.36){\line(1,0){.07}}
    \put(0,-0.35){\line(1,0){.04}}
  }

  \linethickness{2.5mm}
  \color{red}
  \put(0.07,-0.22){\line(0,-1){.06}}
  \put(0.07,-0.28){\line(-1,1){.03}}\put(0.07,-0.28){\line(1,1){.03}}
  \put(0.07,-0.21){\small Large Sell Order {\tiny (B1)}}
  \linethickness{1.5mm}
  \color{mgrey}
  \put(0.00,-0.28){\line(1,0){.20}}

  \linethickness{2.5mm}
  \color{blue}
  \put(0.41,-0.2){\line(0,-1){.05}}
  \put(0.41,-0.2){\line(-1,-1){.03}}\put(0.41,-0.2){\line(1,-1){.03}}
  \put(0.4,-0.28){\small Large Buy Order {\tiny (B2)}}
  \linethickness{1.5mm}
  \color{mgrey}
  \put(0.34,-0.20){\line(1,0){.20}}

\end{picture}


\caption{\rred{Idealized} kinematics of market impact caused by bad synchronization (A1--A2--A3 sequence) 
and preservation of the market depth thanks to a market maker agreeing to support market risk (B1--B2-B3 sequence).}
  \label{fig:syncorders}
\end{figure}

\item \emph{Post-fragmented markets}: regulations \rred{have} evolved \rred{with the aim of} implementing more competition 
across each layer of \rred{Figure \ref{fig:syncorders}} (especially across market operators) and \rred{increasing} transparency:
  \begin{itemize}
  \item in the US, Reg NMS decided to keep the competition inside the layer of market operators: it \rred{requires}
an Exchange or an Electronic Communication Network (ECN) to route an order to the platform that offers the best match 
(it is called the \emph{trade-through rule}). For instance, if a trader sends a buy order at \$10.00 to BATS where 
the best ask price is \$9.75 and if the best ask for this stock is \$9.50 on NYSE, BATS has to re-route the order to NYSE. 
This regulation needs two important elements: 
\begin{enumerate}
\item[(1)] a way of pushing to all market operators the best bid and ask of any
available market with accuracy (it raises concerns linked to the latency of market data); 
\item[(2)] that buying at \$9.50 on 
NYSE is always better for a trader than buying at \$9.75 on BATS, meaning that the other trading costs (especially 
clearing and settlement costs) are the same. 
\end{enumerate}
The data conveying all the best bid and asks is called 
the \emph{consolidated pre-trade tape} and its best bid and offer is called the \emph{National Best-Bid and Offer} (NBBO).
\item in Europe, mainly because of the diversity of the clearing and settlement channels, MiFID 
\rred{allows the competition to be extended} to the intermediaries: they are in charge of defining
 their \emph{Execution Policies} describing how and why they will route and split orders across market operators. 
The European Commission thus relies on competition between execution policies 
\rred{as the means of selecting} the best 
way of splitting orders, taking into account all trading costs. As a consequence, Europe does not
 have any officially consolidated pre-trade tape.
  \end{itemize}

\begin{figure}[!h]
  \centering
  \includegraphics[width=\textwidth]{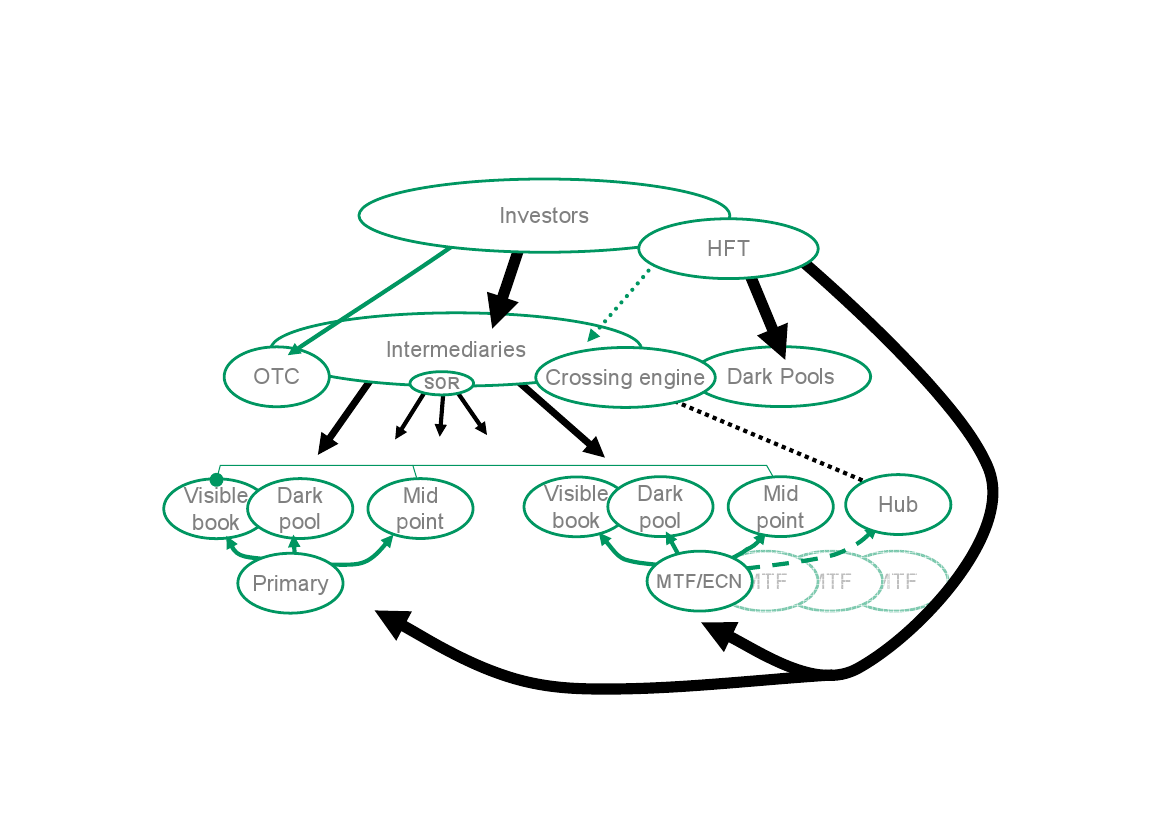}  
  \caption{\rred{Idealized} post-fragmentation market microstructure.}
  \label{fig:MMSfrag}
\end{figure}

  Despite these differences, European and US electronic markets have a lot in common: their microstructures 
evolved similarly to a state where latency is crucial and \emph{High Frequency Market-Makers} 
(also called \emph{High Frequency Traders}) became the main liquidity providers of the market.

   Figure \ref{fig:MMSfrag} gives \rred{an idealized} view of this fragmented microstructure:
  \begin{itemize}
  \item A specific class of investors: the \emph{High Frequency Traders} (HFT) \rred{are}
an essential part of the market; \rred{by} investing more than other market participants in technology, 
thus reducing their latency to markets, they \rred{have} succeeded in:
\begin{itemize}
\item implementing market-making-like behaviors at high frequency;
\item providing liquidity at the bid and ask prices when the market has \rred{low probability}
of moving (thanks to statistical models);
\item \rred{being} able to cancel very \rred{quickly} resting orders \rred{in order} 
to minimize the market risk exposure of their inventory;
\end{itemize} they are said to be \rred{feature in} 70\% of the transactions in US Equity markets, 
40\% in Europe and 30\% in Japan in 2010. Their interactions with the market have been intensively studied by 
 \cite{citeulike:8423311}.
  \item Because they are the main customers of market operators, HFTs \rred{offered} new features 
\rred{making it easier to conduct their business}: 
low latency access to matching engines (better quality of service and \emph{co-hosting}; 
i.e. the ability to locate their computers physically close to the ones of the matching engines), 
and even \emph{flash orders} (\rred{knowing} before other market participants that 
an order is being inserted in the order-book).
  \item Market participants \rred{that were} not proprietary high-frequency traders \rred{also}\break \rred{sought} specific
 features of the order books, mainly to hide their interests \rred{from} high frequency traders: 
\emph{Dark Pools}, implementing anonymous auctions (i.e. partially observable), are part of this offer.
  \item \rred{The} number of market operators as firms does not increase that much
 when a market goes from non-fragmented to fragmented, because of high technological costs 
linked to a fragmented microstructure. On the other hand, each operator offers more products 
(order books) to clients when fragmentation increases. The BATS and Chi-X Europe merged and 
the London Stock Exchange--Milan Stock Market--Turquoise trading \rred{also formed a single group}.
 Looking at the European order-books offered by NYSE-Euronext \rred{in 2011 only}, we have:
    \begin{itemize}
    \item several visible (i.e. \emph{Lit}) \emph{order books}: one for Paris--Amsterdam--Brussels stocks, 
another (NYSE--Arca Europe) for other European names;
    \item \emph{Mid-points}: an order book with only one queue \emph{pegged} at the mid-price of a 
reference market (SmartPool);
    \item \emph{Dark pools}: an anonymous order book (i.e. market participants can send 
orders \rred{as} in a Lit book, but no-one can read the state of the book);
    \item \emph{Fixing auctions}, opening and closing the continuous auctions on visible books.
    \end{itemize}
  \end{itemize}
  The result is an interconnected network of liquidity in which each market participant is no 
\rred{longer} located in one layer only: HFTs are \rred{simultaneously} investors and also very 
close to market operators, intermediaries are offering \emph{Smart Order Routers} to split 
optimally orders across all available trading pools \rred{whilst} taking into account the specific liquidity 
needs of each investor. \rred{Thus, market operators 
are close to technology providers}.
\end{itemize}

\begin{figure}[!h]
  \centering
  \includegraphics[width=0.9\textwidth]{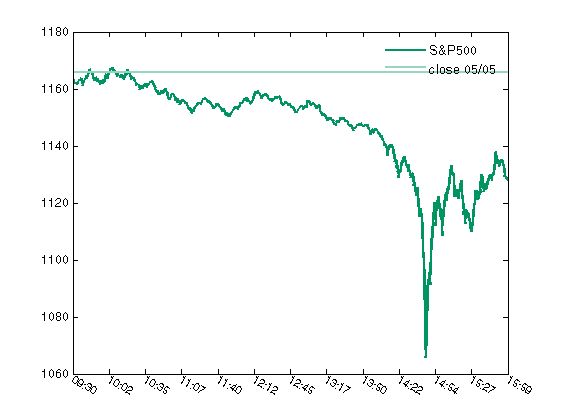}
  \caption{The `Flash Crash': 6 May 2010, US market rapid down-and-up move by almost 
10\% was only due to market microstructure effects.}
  \label{fig:6may}
\end{figure}

\begin{figure}[!h]
  \centering
   \includegraphics[width=0.9\textwidth]{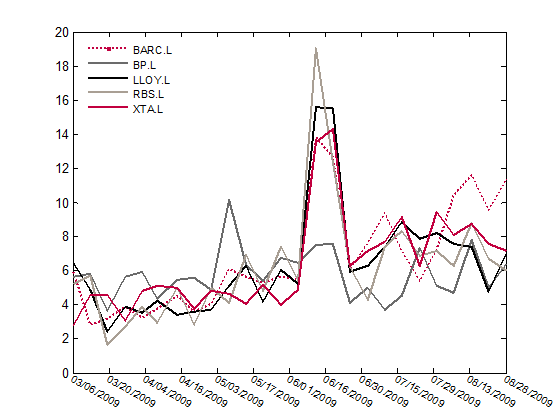}
  \caption{The `Tick war' in June 2009, in Europe. The increase of market share of 
Turquoise (an European Multilateral Trading Facility; MTF) on five Stocks listed on the London Stock Exchange 
following a decrease of the tick size. When other MTFs lowered the tick size, 
the market share \rred{returned} to the previous level.}
  \label{fig:tickw}
\end{figure}

The regulatory task is thus more sophisticated in a fragmented market rather than in a concentrated one:
\begin{itemize}
\item the \emph{Flash Crash} of 6 May 2010 \rred{in} US markets raised concerns about the stability of 
such a microstructure (see Figure~\ref{fig:6may});
\item the cost of surveillance of trading flows across a complex network is higher than in a concentrated one.
\end{itemize}
Moreover, elements of the market design play \rred{many} different roles: the \emph{tick size} for instance, 
is not only the minimum between 
two consecutive different prices, \rred{i.e.,} a constraint on the bid-ask spread, 
it is \rred{also} a key in the competition between market operators. In June 2009, 
European market operators tried to gain market 
shares \rred{by} reducing the tick size on their order books. 
Each time one of them offered a lower tick than others, 
it gained around 10\% of market shares (see Figure \ref{fig:tickw}). 
After \rred{a} few weeks of competition on the tick, they limited this kind of infinitesimal 
\rred{decimation} of the tick thanks 
to a gentleman's agreement obtained under the umbrella of the FESE (Federation of European Security Exchanges): 
such a \rred{decimation had been expensive} in CPU and memory demand for their matching engines.

\paragraph{\rred{An idealized} view of the `Flash Crash'.}
The flash crash   \rred{was accurately} described in \cite{citeulike:8676220}. 
The sequence of events that \rred{led} to a negative jump \rred{in} price and a huge increase \rred{in} 
traded volumes in few minutes, followed by a return to normal in less than 20 minutes can be 
\rred{pictured as follows}:
\begin{enumerate}
\item[(1)] A final investor decided to sell a large amount $v^*$ of shares of 
the E-Mini future contracts, \rred{asking} a broker to take care of this sell 
by electronic means on his behalf.
\item[(2)] The broker decided to use a \emph{PVOL} (i.e. Percentage of Volume) algo, with the 
instruction to follow almost uniformly 9\% of the market volume without regard \rred{to} price or time. 
This participation rate is not uncommon (it is usual to see PVOL algos with the instruction to follow 20\% 
of the market volume).
\item[(3)] The trading algorithm \rred{could} be seen as a trade scheduler splitting the order in slices 
of \rred{one-minute intervals}, expecting to see a traded volume $V_t$ during the $t$th slice 
(meaning that $\mathbb{E}(V_t)\simeq \overline{V}/500$, where $\overline{V}$ is the expected daily traded volume).
\item[(4)] For its first step, the algo began to sell on the future market around 
$v_0=\mathbb{E}(V_0) \times 9/(100-9)\simeq \overline{V}/500\times 0.09$ shares,
\item[(5)] The main buyers of these shares had been intra-day market makers; say that they bought $(1-q)$ of them.
\item[(6)] Because the volatility was quite high on 6 May 2010, the market makers did not feel comfortable 
with such an imbalanced inventory, \rred{and so} decided to hedge it on the cash market, 
selling $(1-q)\times v_t$ shares of a properly weighted basket of equities.
\item[(7)] Unfortunately the buyers of most of these shares (say $(1-q)$ of them again)
were intra-day market makers themselves, who decided \rred{in} their turn to hedge their risk on the future market.
\item[(8)] It immediately increased the traded volume on the future market by $(1-q)^2 v_0$ shares.
\item[(9)] Assuming that intra-day market makers could play this \emph{hot potato game} (as it \rred{was called}
in the SEC--CFTC report), $N$ times in 1 minute, the volume traded on the future market \rred{became} 
$\sum_{n\leq N} (1-q)^{2n} v_0$ larger than expected by the brokerage algo.
\item[(10)] Back to step (4) at $t+1$, the PVOL algo is now late by 
$\sum_{n\leq N} (1-q)^{2n} v_0 \times 8/(100-8)$, and has to sell 
$\overline{V}/500  \times 8/(100-8)$ again; i.e. selling 
$$v_{t+1}\simeq\left( N\times v_t + \frac{\overline{V}}{500} \right)\times 0.08.$$
\end{enumerate}

\begin{figure}[!h]
  \centering
  \includegraphics[width=0.9\textwidth]{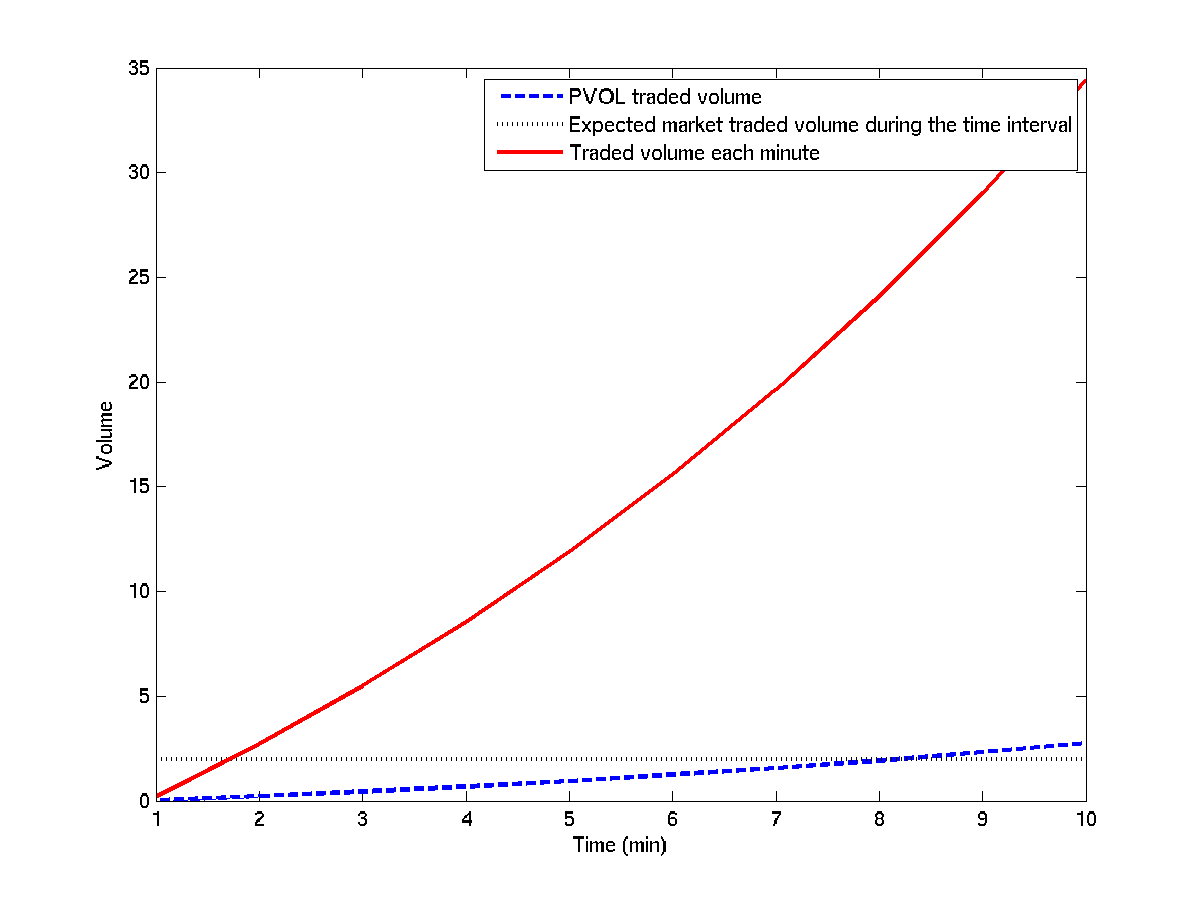}
  \caption{Traded volume of the future market according to the simple \rred{idealized} model with 
$\overline{V}=100$, $T=10$ and $N=2$.}
  \label{fig:modFC}
\end{figure}

Figure \ref{fig:modFC} shows how explosive \rred{the \emph{hot potato game} between 
intra-day market makers can be}, even with \rred{not that high a} frequency trading rate (here $N=1.1$). 
Most of this trading flow \rred{was a selling flow, pushing most  US prices} to very low levels. 
For instance Procter and Gamble quoted from \$60 to a low of \$39.37 in approximately 3.5 minutes.

In reality other effects contributed to the flash crash:
\begin{itemize}
\item only \rred{a few} trading pools implemented circuit breakers that \rred{ought to have frozen} 
the matching engines in case of sudden liquidity event;
\item most market participants only looked at the \emph{consolidated tape} for market data, 
preventing them \rred{noticing} that \rred{trading was frozen on some pools};
\item in the US, most retail flow is internalized by market makers. \rred{At} one point in the day 
these intermediaries decided to hedge their positions on the market \rred{on} their turn, 
 \rred{further affecting the prices}.
\end{itemize}

This glitch in the electronic structure of markets is not an isolated case, even if it \rred{was} 
the largest one. The \rred{combination} of a failure in each layer of the market 
(an issuer of large institutional trades, a broker, HF market-makers, market operators) 
with a highly uncertain market context is \rred{surely} a crucial element of this crash. 
It has moreover shown that most orders do \rred{indeed} reach the order books only through electronic means.

European markets did not suffer from such \emph{flash crashes}, but they have not seen many months 
in 2011 without an outage of a matching engine.

\begin{figure}[!h]
  \centering
  \includegraphics[width=0.9\textwidth]{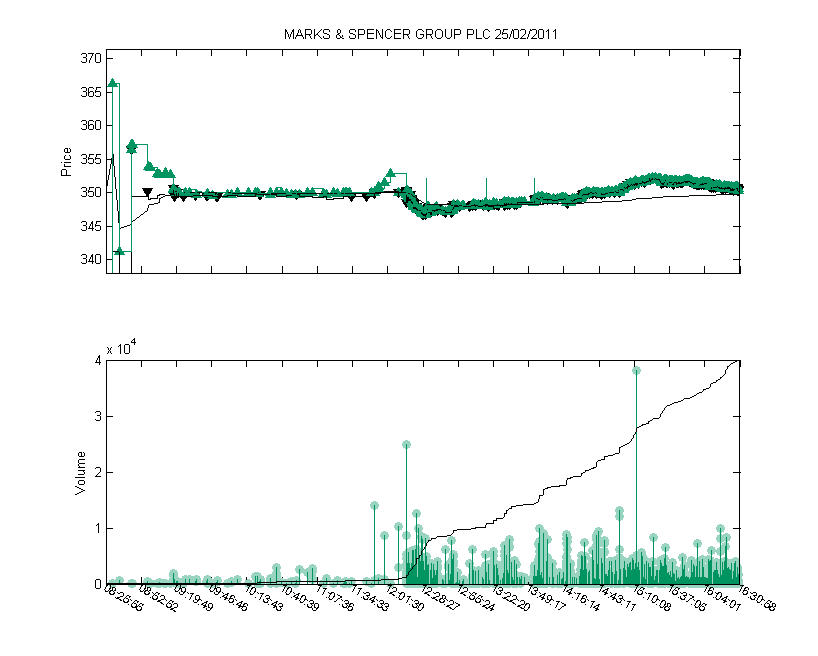}
  \caption{Examples of outages in European equity markets on 25 February 2011. 
The price (top) and the volumes (bottom) when the primary market \rred{opened only} after 12:15 (London time). 
The price did not move \rred{much}.}
  \label{fig:Eoutages}
\end{figure}

\paragraph{European outages.}
Outages are `simply' bugs in matching engines. In such cases, the matching engines of 
one or more trading facilities can be frozen, or just stop \rred{publishing} market data, 
becoming true \emph{Dark Pools}. From a scientific viewpoint, and because in Europe there is no 
\emph{consolidated pre-trade tape} (i.e. each member of the trading facilities needs to 
build by himself
his \emph{consolidated view} of the current European best bid and offer), 
they \rred{can} provide examples of behavior of market participants when they do not all share the 
same level of information \rred{about} the state of the offer and demand.

For instance:
\begin{itemize}
\item when no information is available on primary markets but trading remains open: 
two price formation processes can take place in parallel, one for market participants having 
access to other pools, and the other for participants who just looked at the primary market;
\item (Figure \ref{fig:Eoutages}) when the primary market does not start trading at the very 
beginning of the day: the price does not really move on alternative markets; no `real' 
price formation process takes place during such European outages.
\end{itemize}

The flash crash in US and the European outages emphasizes the \emph{role of information in the 
price formation process}. When market participants are confident that they have access to 
a reliable source of information (during the flash or during some European outages), 
they continue to \emph{mimic} a price formation process which output can be far from efficient. 
\rred{By contrast}, if they do not believe in the information they have, they just freeze their 
price, \rred{observe} behavior and trade at the \emph{last confident price}, \rred{while}
waiting for reliable updates.

\section{Forward and Backward Components of the Price Formation Process}\label{lehalle_sec3}

\begin{figure}[!h]
  \centering
   \includegraphics[width=0.9\textwidth]{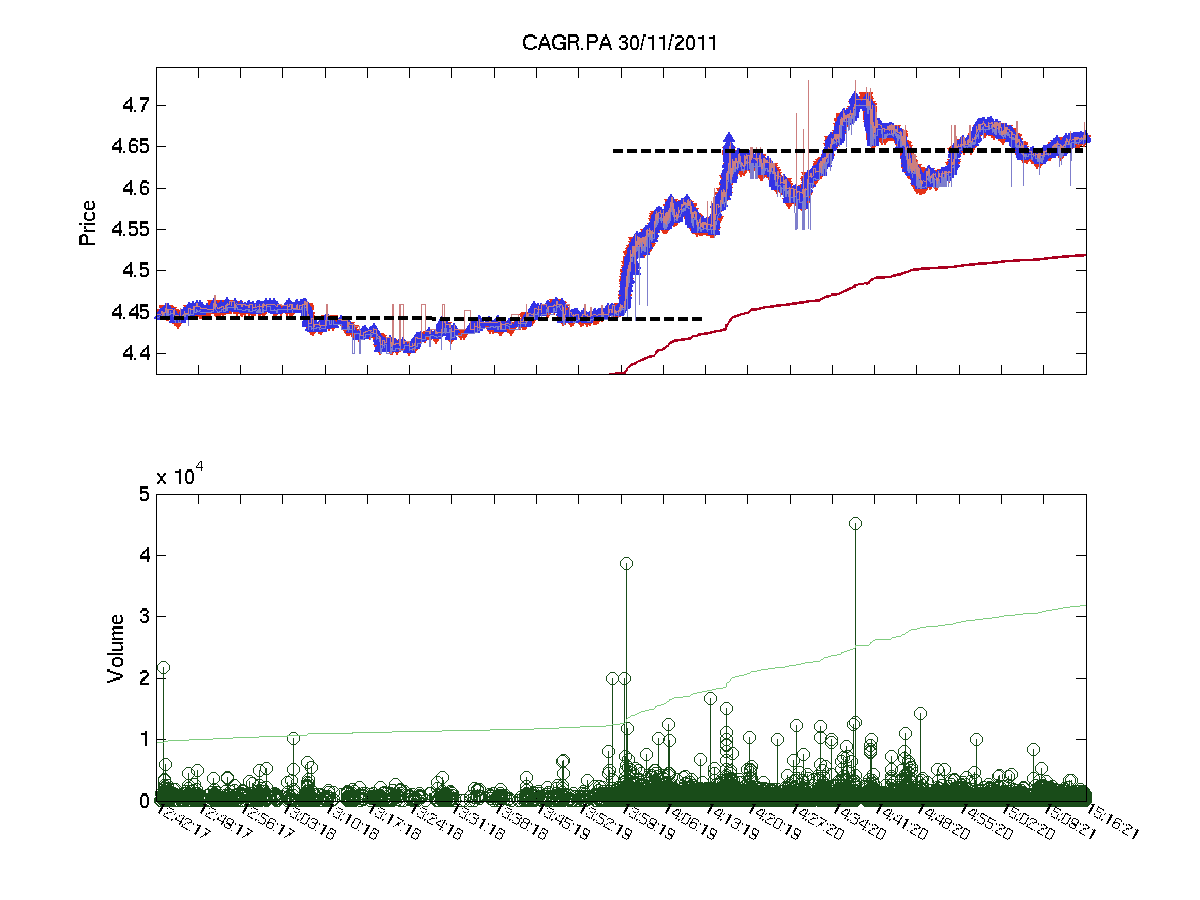}
  \caption{A typical \emph{Price Discovery} exercise: the 30th of November, 2011 on the Cr\'edit Agricole share price (French market). The two stable states of the price are materialized using two dark dotted lines, one before and the other after the announcement by major European central banks a coordinated action to provide liquidity.}
  \label{fig:pdiscovery}
\end{figure}

The literature on market microstructure can be split in two generic subsets:
\begin{itemize}
\item papers with a \emph{Price Discovery} viewpoint, in which the market participants 
are injecting into the order book their views on a fair price. In these papers 
(see for instance \cite{RePEc:ide:wpaper:825,RePEc:eee:jfinec:v:9:y:1981:i:1:p:47-73,citeulike:7604491}),
 the \emph{fair price} is assumed to exist for fundamental reasons (at least in the mind of investors) 
and the order books are implementing a Brownian-bridge-like trajectory targeting this evolving fair price. 
This is a \emph{backward} view of the price dynamics: the investors are updating assumptions on 
the future value of tradeable instruments, and send orders in the electronic order books according 
to the distance between the current state of the offer and demand and this value, driving the quoted 
price to \rred{some average of what they expect}.

Figure \ref{fig:pdiscovery} shows a price discovery pattern: the price of the stock 
changes for fundamental reasons, and the order book dynamics react accordingly generating 
more volume, more volatility, and a price jump.
\end{itemize}

\begin{itemize}
\item Other papers rely on a \emph{Price Formation Process} viewpoint. For their authors 
(most of them econophysicists, see for instance \cite{farmer03a, citeulike:1618840} 
or \cite{citeulike:5823204} for a review of agent based models of order books) 
the order books are \emph{building the price} in a forward way. The market participants take 
decisions with respect to the current orders in the books making assumptions of the future value 
of their inventory; it is a \emph{forward} process.
\end{itemize}

Following \cite{citeulike:7621540}, \rred{one can try to crudely model} these two dynamics simultaneously. 
In a framework with an infinity of agents (using a Mean Field Game approach, 
see \cite{citeulike:3614137} for more details), 
the order book at the bid (respectively at the ask), is a density $m_B(t,p)$ 
(resp. $m_A(t,p)$) of agents agreeing at time $t$ to buy (resp. sell) at price $p$. 
In such a continuous framework, there is no bid--ask spread and the \emph{trading price} 
$p^*(t)$ is such that there is no offer at a price lower than $p^*(t)$ 
(and no demand at a price greater then $p^*(t)$).
\rred{Assuming diffusivity}, the two sides of the order book are \rred{subject to 
the following simple partial differential equations}:
\begin{eqnarray*}
\partial_{t}m_B\left(t,p\right)-\frac{\varepsilon^{2}}{2}\partial_{pp}^{2}m_B(t,p)&=&\lambda(t)\delta_{p=p^*(t)}\\
\partial_{t}m_A\left(t,p\right)-\frac{\varepsilon^{2}}{2}\partial_{pp}^{2}m_A(t,p)&=&\lambda(t)\delta_{p=p^*(t)}.
\end{eqnarray*}
\rred{Moreover}, the trading flow at $p^*(t)$ is clearly defined as
 $$\lambda(t) = -\frac{\varepsilon^{2}}{2}\partial_{p}m_B\left(t,p^*(t)\right) = 
\frac{\varepsilon^{2}}{2}\partial_{p}m_A\left(t,p^*(t)\right). $$

It is then possible to define a regular order book $m$ joining the \rred{bid and ask sides} by
$$m(t,p) = \left\lbrace
   \begin{array}{r c l}
      m_B(t,p)  &, \quad& \mbox{\rm if} \ p \le p^*(t) \\
      -m_A(t,p) & , \quad & \mbox{\rm if} \ p > p^*(t) \\
   \end{array}
   \right.$$
which satisfies a \rred{single} parabolic equation:
\begin{equation}
  \label{eq:macro:3}
  \partial_{t}m\left(t,p\right)-\frac{\varepsilon^{2}}{2}\partial_{pp}^{2}m(t,p)=
-\frac{\varepsilon^{2}}{2}\partial_{p}m\left(t,p^*(t)\right)
\left(\delta_{p=p^*(t)-a}-\delta_{p=p^*(t)+a}\right)
\end{equation}
with a limit \rred{condition} $m(0,\cdot)$ given on the domain $[p_{\min},p_{\max}]$ and, 
for instance, Neumann conditions at $p_{\min}$ and $p_{\max}$.

Such a \emph{forward process} describes the order book dynamics without any impact \rred{on} investors' 
fundamental views (it is a \emph{price formation process} model).

\begin{figure}[!h]
  \centering
  \includegraphics[width=\textwidth]{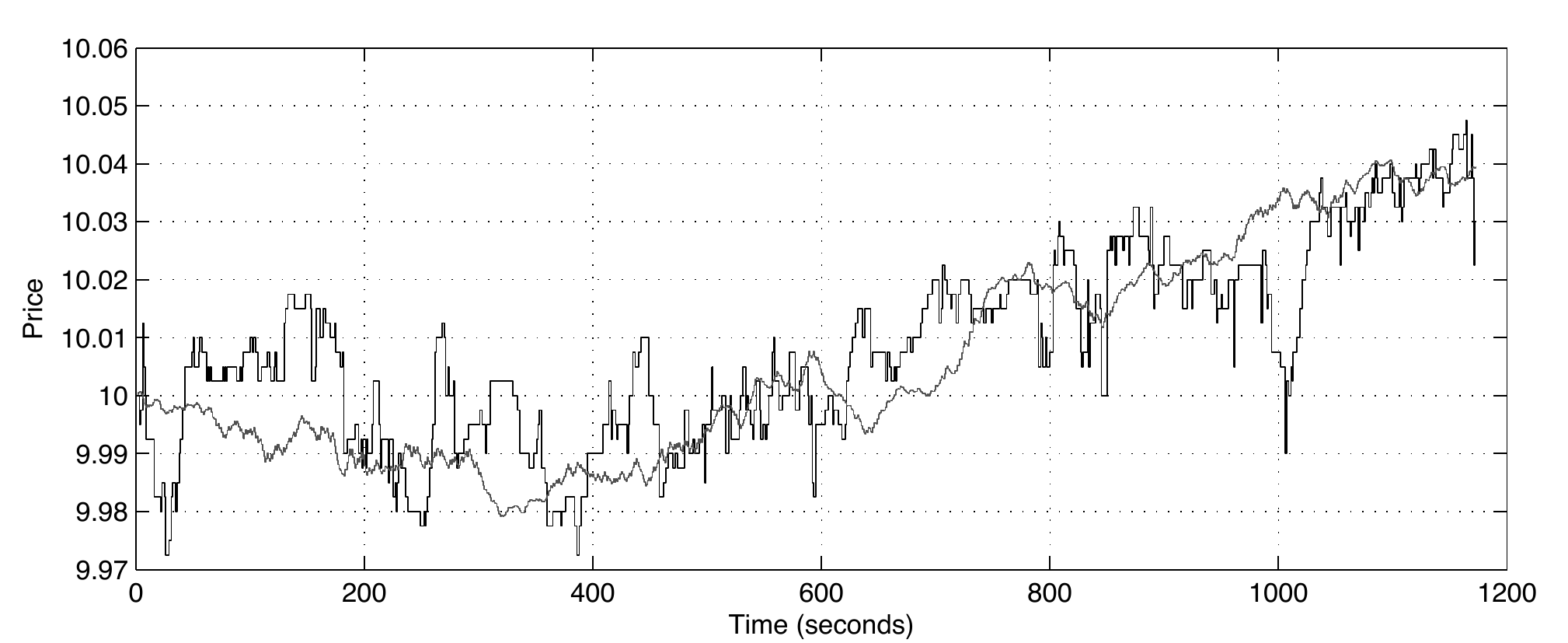}
  \caption{Simulation of the dynamics \rred{modeling an order book using} a forward--backward 
approach: the `fair price' is the continuous grey line and the realized price is the stepwise dark one.}
  \label{fig:mfg}
\end{figure}

\rred{Lehalle et al.}  then introduce a more complex source to re-inject the orders in 
\rred{books containing market participants' forward views} on the price. 
For instance, a trend follower with a time horizon of $h$ buying at price $p^*(t)$ at 
time $t$ \rred{aims} to unwind \rred{his} position at a higher (i.e. `trend targeted') 
price and thus insert an order in the book accordingly (around $p^*(t)+(p^*(t)-p^*(t-h))$: see 
the paper for more details). Figure \ref{fig:mfg} shows an example of such a dynamic.

This is a way of introducing investor-driven views \rred{into} the model, 
which are essentially \emph{backward}: a trend follower \rred{agrees to be} part of a transaction 
because he believes that the price will continue to move in the same direction \rred{over} 
his investment time scale. This future price of the share is at the root of his decision. 
This is an injection of a \emph{price discovery} component in the model.

\section{From Statistically Optimal Trade Scheduling to Microscopic Optimization of Order Flows}\label{lehalle_sec4}

Modeling the price formation dynamics is of interest for \rred{both regulators and  policy makers}. 
It enables them to understand the potential effects of a regulatory or rule change on the efficiency 
of the whole market (see for instance \cite{FOU06} for an analysis of the introduction of 
competition among trading venues on the efficiency of the markets). It thus helps in understanding 
potential links between market design and systemic risk.

In terms of risk management inside a firm hosting trading activities, it is more important to 
understand the trading cost of a position, which can be understood as its \emph{liquidation risk}.

From the viewpoint of one trader \rred{versus} the whole market, 
three key phenomena have to be controlled:
\begin{itemize}
\item the \emph{market impact} (see \cite{citeulike:3320208,NAT03,citeulike:4368376,citeulike:4325901,Bouchaud06}) 
\rred{which} is the market move generated by selling or buying a large amount of shares 
(all else being equal); it comes from the forward component of the price formation process, and can be temporary
 if other market participants (\rred{they are} part of the backward component of the price discovery dynamics) 
provide enough liquidity to the market to bring back the price \rred{to} its previous level;
\item \emph{adverse selection}, capturing the fact that providing too much (passive) 
liquidity via limit orders \rred{enables} the trader \rred{to} maintain the price at an artificial level; 
not a lot of literature is available \rred{about} this effect, which has been nevertheless identified 
by practitioners \citep{citeulike:6716078};
\item and the uncertainty on the fair value of the stock that can move the price during the trading process; 
it is often referred as the \emph{intra-day market risk}.
\end{itemize}

\subsection{Replacing market impact by statistical costs}

A \rred{framework now widely used for controling} the overall costs of the liquidation of a portfolio 
was proposed by Almgren and Chriss in the late 1990s \cite{OPTEXECAC00}. Applied to 
\rred{the trade of a} single stock, this framework:
\begin{itemize}
\item cuts the trading period into an arbitrary number of intervals $N$ of a chosen duration $\delta t$,
\item models the \emph{fair price} moves thanks to a Gaussian random walk:
  \begin{equation}
    \label{eq:9}
      S_{n+1}=S_n + \sigma_{n+1}\sqrt{\delta t} \; \xi_{n+1}
  \end{equation}
\item models the \emph{temporary market impact} $\eta_n$ inside each time bin using a 
power law of the trading rate (i.e. the ratio of the traded shares $v_n$ by the trader over 
the market traded volume during the same period $V_n$):
  \begin{equation}
    \label{eq:2}
    \eta(v_n)=a\,\psi_n+\kappa\, \sigma_n\sqrt{\delta t} \left(\frac{v_n}{V_n}\right)^\gamma
  \end{equation}
where $a$, $\kappa$ and $\gamma$ are parameters, and $\psi$ is the half bid-ask spread;
\item \rred{assumes  the \emph{permanent market impact} is  linear} in the participation rate;
\item uses a mean--variance criterion and minimizes it to obtain the optimal sequence of shares to buy 
(or sell) through time.
\end{itemize}

It is \rred{important first} to notice that there is an implicit relationship 
between the time interval $\delta t$ and the temporary market impact function: 
without changing $\eta$ and simply by choosing a different \rred{time slice}, the cost 
of trading \rred{can be} changed. It is in fact not possible to choose $(a,\kappa,\gamma)$ and 
$\delta t$ independently; they have to be chosen \rred{according} to the decay of the market 
impact on the stock, provided that most of the impact is kept in a time bin of 
size $\delta t$. Not all the decay functions are compatible with this view 
(see \cite{citeulike:10363463} for details about available market impact 
models and their interactions with trading). Up to now the terms in $\sqrt{\delta t}$ have 
been ignored. \rred{Note also that the parameters $(a,\kappa,\gamma)$ are relevant at this time scale}.

\rred{One should not regard this framework as if it were based on structural model assumptions} 
(i.e. that the market impact really has this shape, or that the price moves really are Brownian), 
\rred{rather, as if it were a statistical one}.
With such a viewpoint, any practitioner can use the database of its past executed orders and 
perform an econometric study of its `trading costs' on any interval, $\delta t$,
of time (see \cite{citeulike:4368376} for an analysis of this kind on the whole duration of the order). 
If a given time scale succeeds \rred{in capturing, with enough accuracy, 
the parameters of a trading cost model, then that model can be used to optimize trading. 
Formally, the result of such a statistical approach would be the 
same as that of a structural one, as we will show below. But it is possible to go one step further, 
and to take into account the statistical properties of the variables (and parameters) of interest}.

Going back to the simple case of the liquidation of one stock without any permanent market impact, 
the value (which is a random variable) of a buy of $v^*$ shares in $N$ bins of size 
$v_1, v_2,\ldots, v_N$ \rred{is}
\begin{eqnarray}
  \nonumber
  W(v_1, v_2,\ldots, v_N) %
    &=& \sum_{n=1}^N v_n ( S_n +  \eta_n(v_n) )\\
\nonumber
  &=& S_0 v^* + %
  \underbrace{\sum_{n=1}^N \sigma_n \xi_n x_n}_{\mbox{market move}}  \\
  \label{eq:1}
  &&  %
  \qquad +\underbrace{\sum_{n=1}^N a\, \psi_n  (x_n-x_{n+1}) + \kappa \,\frac{\sigma_n}{V_n^\gamma} \, 
(x_n-x_{n+1})^{\gamma+1}}_{\mbox{market impact}},\qquad~
\end{eqnarray}
using the \emph{remaining quantity to buy}: \rred{that is,} $x_n=\sum_{k\geq n} v_k$ 
instead of the instantaneous volumes $v_n$. To obtain \rred{an answer in as closed a form} as possible, 
$\gamma$ will be taken equal to 1 (i.e. linear market impact).
\rred{(See \cite{citeulike:5797837} for a more sophisticated model 
and more generic utility functions rather than the idealized model which we adopt here in order to obtain clearer 
illustrations of phenomena of interest.)} 

To add a practitioner-oriented flavor to our upcoming optimization problems, just introduce a set of independent 
random variables $(A_n)_{1\leq n\leq N}$ to model the \emph{arbitrage opportunities} during time slices. 
It will reflect \rred{our expectation} that the trader will be able to buy shares at price $S_n-A_n$ 
during slice $n$ rather than at price $S_n$.

Such an effect can be used to inject a statistical arbitrage approach into optimal trading or to take into 
account the \rred{possibility of crossing} orders at mid price in Dark Pools or Broker Crossing Networks 
(meaning that the expected trading costs should be smaller during given time slices). 
Now the cost \rred{of buying} $v^*$ shares is:
\begin{eqnarray}
  \label{eq:4}
  W(\mathbf{v}) &=& S_0 v^* + \sum_{n=1}^N \sigma_n \xi_n x_n +  
\sum_{n=1}^N (a\, \psi_n - A_n)  v_n +  \kappa \, \frac{\sigma_n}{V_n} \, v_n^2
\end{eqnarray}

\paragraph{Conditioned expectation optimization.}
The expectation of this cost, 
$$\mathbb{E}(W\vert (V_n,\sigma_n,\psi_n)_{1\leq n\leq N}),$$ 
given the market state, \rred{can be written as}
\begin{equation}
  \label{eq:3}
  C_0= %
  S_0 v^* +  \sum_{n=1}^N (a\, \psi_n - \mathbb{E} A_n)  v_n +  \kappa \, \frac{\sigma_n}{V_n} \, v_n^2 ,
\end{equation}
\rred{A} simple optimization under constraint (to ensure $\sum_{n=1}^N v_n=v^*$) gives
\begin{equation}
  \label{eq:5}
  v_n = w_n \left( v^* + \frac{1}{\kappa}\left( \left(\mathbb{E} A_n -\sum_{\ell=1}^N w_\ell \mathbb{E} A_\ell\right) 
-a \left( \psi_n -\sum_{\ell=1}^N w_\ell\psi_\ell \right) \right) \right),
\end{equation}
where $w_n$ are weights proportional to the inverse of the market impact factor: 
$$w_n=\frac{V_n}{\sigma_n}\left(\sum_{\ell=1}^N \frac{V_\ell}{\sigma_\ell}\right)^{-1}.$$

Simple effects can be deduced from this first \rred{idealization}.
\begin{enumerate}
\item[(1)]  Without any arbitrage opportunity and without any bid-ask cost 
(i.e. $\mathbb{E} A_n=0$ for any $n$ and $a=0$), the optimal trading rate is proportional to the 
inverse of the market impact coefficient: $v_n=w_n\cdot v^*$.
Moreover, when the market impact has no intra-day seasonality, $w_n=1/N$ implying that the optimal trading rate is linear.

\item[(2)] Following formula (\ref{eq:5}) it can be seen that the \rred{greater} 
the expected arbitrage gain (or the lower the spread cost) on a slice compared 
to the market-impact-weighted expected arbitrage gain (or spread cost) over the \rred{full trading} interval, 
the \rred{larger the} quantity to trade during this slice.
More quantitatively:
$$\df{v_n}{\mathbb{E} A_n}=\frac{w_n}{2\kappa} (1-w_n) >0,\; \df{v_n}{\psi_n}=-\frac{a}{2\kappa}(1-w_n)w_n<0.$$

This result gives the \emph{adequate weight} \rred{for applying}
to the expected arbitrage gain \rred{in order} to translate it \rred{into} an adequate trading rate \rred{so as to profit} 
on arbitrage opportunities on average. Just note that usually the expected arbitrage gains increase with market 
volatility, \rred{so} the $w_n$-weighting is consequently of interest to balance this effect optimally.
\end{enumerate}

\paragraph{Conditioned mean-variance optimization.}
Going back to a mean--variance optimization of the cost \rred{of  buying} progressively $v^*$ shares, the criterion 
for minimizing (using a risk aversion parameter $\lambda$) \rred{becomes}
\begin{eqnarray}
\label{eq:6} 
C_\lambda &=& \mathbb{E}(W\vert (V_n,\sigma_n,\psi_n)_{1\leq n\leq N}) + 
\lambda \mathbb{V}(W\vert (V_n,\sigma_n,\psi_n)_{1\leq n\leq N})\nonumber\\
&=& S_0 v^*  + \sum_{n=1}^N (a \psi_n-\mathbb{E} A_n) \X(n) + 
\left(\kappa\frac{\sigma_n}{V_n}+\lambda \mathbb{V} A_n\right) \X(n)^2 + \lambda \sigma_n^2 x_n^2.\qquad~
\end{eqnarray}

To minimize $C_\lambda$ \rred{when it is} only constrained by terminal conditions on $x$ 
(i.e. $x_0=v^*$ and $v_{N+1}=0$), it is enough to cancel its derivatives with respect to any 
$x_n$, leading to \rred{the} recurrence relation
\begin{eqnarray}
\nonumber  \left( \frac{\sigma_n}{V_n}+\frac{\lambda}{\kappa} \mathbb{V} A_n\right) x_{n+1} &=& %
  \frac{1}{2\kappa}(a ( \psi_{n-1}-\psi_n) - (\mathbb{E} A_{n-1}-\mathbb{E} A_n)  )\\
 \nonumber &&\quad  + \left(\frac{\lambda}{\kappa} \sigma_n^2+  \left(\frac{\sigma_n}{V_n}+\frac{\lambda}{\kappa} \mathbb{V} A_n + \frac{\sigma_{n-1}}{V_{n-1}}+\frac{\lambda}{\kappa} \mathbb{V} A_{n-1}\right) \right) x_n  \\
  &&\qquad  - \left( \frac{\sigma_{n-1}}{V_{n-1}}+\frac{\lambda}{\kappa} \mathbb{V} A_{n-1}\right) x_{n-1} .
\end{eqnarray}

\rred{This} shows that the variance of the arbitrage has an  effect similar \rred{to that of} 
the market impact (through a risk-aversion rescaling), and that the risk-aversion parameter 
acts as a multiplicative factor \rred{on} the market impact, meaning that within an arbitrage-free 
and spread-costs-free framework (i.e. $a=0$ and $\mathbb{E} A_n=0$ for all $n$), 
the market impact model \rred{for} any constant $b$ has no effect on the final result as \rred{long} as 
$\lambda$ is replaced by $b\lambda$.

\begin{figure}[!h]
  \centering
  \includegraphics[width=0.9\textwidth]{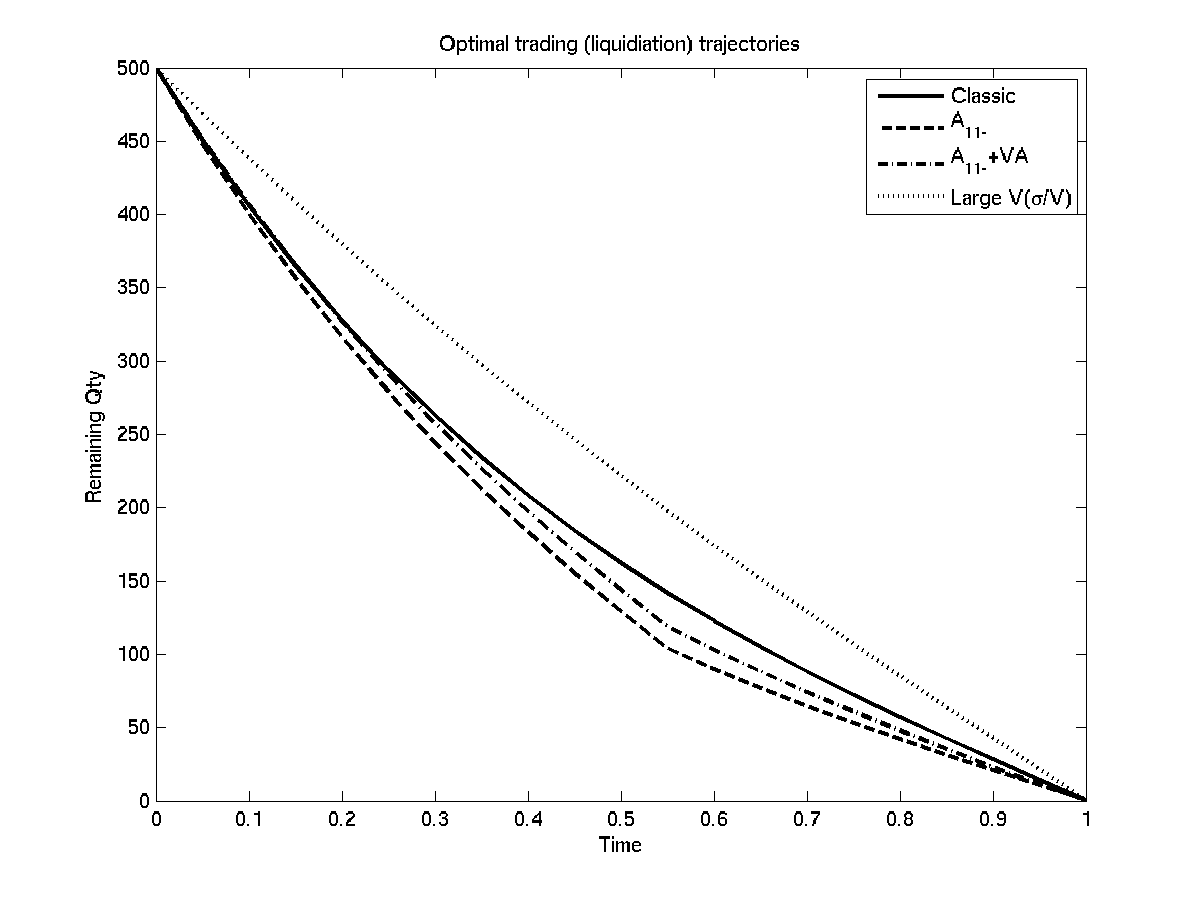}
  \caption{Examples of optimal trading trajectories for mean--variance criteria: the classical result 
(Almgren--Chriss) \rred{is the} solid line, the dotted line is for high variance of the 
variable of interest ($\sigma/V$), the semi-dotted ones for an arbitrage opportunity 
($A_{11+}$ means  after the 11th period; and $A_{11+}+V\! A$ means  adding expected variance 
to the arbitrage opportunity).}
  \label{fig:optrate1}
\end{figure}

Figure \ref{fig:optrate1} compares optimal trajectories coming from different criteria and parameter values.

\paragraph{A statistical viewpoint.}
The two previous examples show how easy it is to include effects in this sliced mean-variance framework.
\rred{The} implicit assumptions are:
\begin{itemize}
\item \rred{within} one time-slice, it is possible to capture the market impact (or \emph{trading costs}) 
using model (\ref{eq:2});
\item the trader \rred{knows} the traded volumes and market volatility in advance.
\end{itemize}

\begin{figure}[!h]
  \centering
  \includegraphics[width=\textwidth]{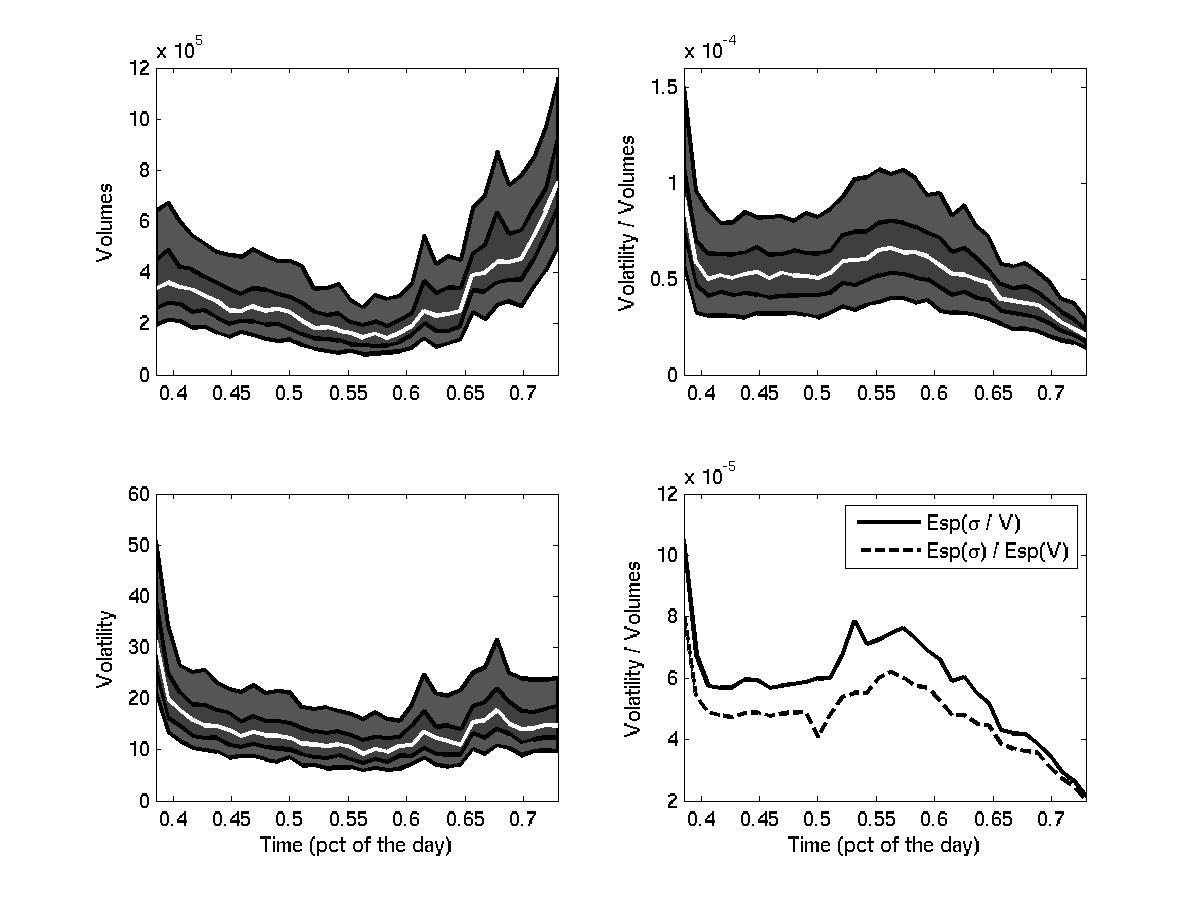}
  \caption{Typical intra-day traded volume (top left) and realized volatility (bottom left) 
profiles  (i.e. intra-day seasonalities on traded volumes and market volatility) 
with their quantiles of level 25\% and 75\%. The $x$-axis is time. The top 
right chart contains the quantiles of the ratio of interest $\sigma/V$. The bottom
right ones shows the difference between the expectation of the ratio (solid line) 
and the ratio of the expectations (dotted line).}
  \label{fig:vcurves}
\end{figure}

\rred{In practical terms}, the two assumptions come from statistical modeling:
\begin{itemize}
\item The market impact parameters $a,\kappa$ and $\gamma$ are estimated on a large database of 
trades using a maximum likelihood or MSE methods; the reality is consequently that the market 
model has the following shape:
  \begin{equation}
    \label{eq:7}
    \eta(v_n)=a\,\psi_n+\kappa\, \sigma_n\sqrt{\delta t} \left(\frac{v_n}{V_n}\right)^\gamma + \varepsilon,
  \end{equation}
where $\varepsilon$ is an i.i.d. noise.
\item Moreover, the market volatility and traded volumes are estimated using historical data and market 
context assumptions (to take into account at least the scheduled news, 
such as the impact of the expiry of derivative products on the volume of the cash market; 
see Figure \ref{fig:vcurves} for typical estimates).
\end{itemize}

Taking these statistical modeling steps into account in the classical mean--vari\-ance 
criterion of (\ref{eq:6}), changes \rred{that equation} into its unconditioned version:
\begin{eqnarray}
  \label{eq:8}
\nonumber  {\tilde C}_\lambda &=& \mathbb{E}(W) + \lambda \mathbb{V}(W)\\
  \nonumber &=& %
   S_0 v^* + \sum_{n=1}^N (a \mathbb{E}\psi_n-\mathbb{E} A_n) \X(n) \\
  \nonumber && %
  \quad + \left(\kappa\,\mathbb{E}\left(\frac{\sigma_n}{V_n}\right)+\lambda (a^2\mathbb{V} \psi_n 
+ \mathbb{V} A_n + \mathbb{V} \varepsilon)\right) \X(n)^2 \\
   && %
  \qquad + \lambda \sigma_n^2 x_n^2 + \lambda \kappa^2\mathbb{V}\left(\frac{\sigma_n}{V_n}\right)\X(n)^4 .
\end{eqnarray}
The \rred{consequences} of using this criterion rather than the conditioned one are clear: 
\begin{itemize}
\item the simple plug-in of empirical averages of volumes and volatility in criterion 
(\ref{eq:6}) instead of the needed expectation of the overall trading costs leads \rred{us}
to use $(\mathbb{E}\sigma_n)/(\mathbb{E} V_n)$ instead of $\mathbb{E}(\sigma_n/V_n)$. 
Figure \ref{fig:vcurves} shows typical differences between the two quantities.

\item If the uncertainty on the market impact is huge (i.e. the $\mathbb{V} \varepsilon$ term dominates all others), 
then the optimal trading strategy is to trade linearly, which is also the solution of a \rred{purely} 
expectation-driven minimization with no specific market behavior linked with time.
\end{itemize}

Within this new statistical trading framework, the inaccuracy of the models and 
the variability of the market context are taken into account: the obtained optimal trajectories will 
no \rred{longer} have to follow sophisticated \rred{paths} if the models are not realistic enough.

Moreover, it is not difficult to solve the optimization program associated to this 
new criterion; the new recurrence equation is a polynomial of degree 3. 
Figure \ref{fig:optrate1} gives illustrations of \rred{the results obtained}.

\rred{Many other effects can be introduced} in the framework, \rred{such as} auto-correlations 
on the volume--volatility \rred{pair}. This statistical framework does not embed recent and 
\rred{worthwhile} proposals such as the decay of market impact \citep{citeulike:10363463} 
or a set of optimal stopping times \rred{that avoid a uniform and a priori sampled} time 
\citep{citeulike:5797837}. It is nevertheless simple enough so that most 
practitioners can use it in order to include their views of the market conditions 
and the efficiency of their interactions with the market \rred{on} a given time scale; 
it can be compared to the Markowitz approach for quantitative portfolio allocation \citep{citeulike:571949}.

\subsection{An order-flow oriented view of optimal execution}

\rred{Though price dynamics in quantitative finance are often modeled using diffusive processes}, 
just looking at prices of transactions in a limit order book convinces \rred{one} that a 
more discrete and event-driven class of model \rred{ought} to be used; 
at a time scale of several minutes or more, the \rred{assumptions of diffusivity} 
used in equation (\ref{eq:9}) to model the price are not that \rred{bad}, 
but even at this scale, the `\emph{bid--ask bounce}' has to be taken into account 
\rred{in order} to be able to estimate with enough accuracy the intra-day volatility. 
The effect on volatility estimates of the rounding of a diffusion \rred{process}
\rred{was} first studied in \cite{VQ96}; \rred{since} then other effects have been taken into account, 
such as an additive \emph{microstructure noise} \citep{scales05}, 
sampling \citep{AITJAC07} or \emph{liquidity thresholding} -- \rred{also known as} uncertainty zones -- 
\citep{citeulike:8317402}. Thanks to \rred{all these models}, 
it is now possible to use high frequency data to estimate the volatility of an underlying 
diffusive process generating the prices without being polluted by the signature plot 
effect (i.e. an explosion of the usual empirical estimates of volatility when high frequency data are used).

Similarly, advances have been made \rred{in  obtaining} accurate estimates of \rred{the} correlations between two underlying 
prices \rred{thereby} avoiding the drawback of the \emph{Epps effect} 
(i.e. a collapse of usual estimates of correlations at small scales \citep{YOSHI05}).

To optimize the interactions of trading strategies with the order-books, 
it is \rred{necessary} to zoom \rred{in} as much as possible and to model 
most known effects taking place at this time scale (see \cite{Bouchaud06,citeulike:1618840}).
Point processes have been \rred{successfully used for this purpose, in particular} 
because they can embed \rred{short-term} memory modeling \citep{citeulike:7012175,citeulike:7012187}.
Hawkes-like processes have most of these interesting properties and exhibit diffusive behavior 
when the time scale is zoomed out \citep{citeulike:7344893}. 
To model the prices of transactions at the bid $N^b_t$ and at the ask $N^a_t$, two 
coupled Hawkes processes can be used. Their intensities $\Lambda^b_t$ and $\Lambda^a_t$ 
are stochastic and are governed by 
$$\Lambda^{a/b}_t = \mu^{a/b} + c\, \int_{\tau<t} e^{ -k (t-\tau)} \, dN^{b/a}_t;$$
\rred{here $\mu^b$ and $\mu^a$ are constants}.
In such a model the more transactions at the bid (resp. ask), the more \rred{likely will there be} one 
at the opposite price in \rred{the near} future.

The next qualitative step \rred{is} to link the prices with the traded volumes. 
It has recently been shown that under some assumptions that are almost \rred{always} true for 
very liquid stocks in a calm market context (a constant bid--ask spread and no dramatic change 
in the dynamics of \rred{liquidity-providing} orders), there is a correspondence between a 
two-dimensional point process of the quantities available at the first limits and the price 
of the corresponding stock \citep{citeulike:8318790,citeulike:8531765}.

To understand the mechanism underlined by such an approach, just notice that the set of stopping 
times defined by the instants when the quantity at the first ask crosses zero 
(i.e. ${\cal T}^a=\{\tau:Q^a_\tau=0\}$) exactly maps the increases of prices 
(if the bid--ask spread is constant). 
Similarly the set ${\cal T}^b=\{\tau:Q^b_\tau=0\}$ maps the decreases of the price.

Despite these valuable proposals for modeling the dynamics of the order book at small time scales, 
\rred{they have not yet been directly used in an optimal trading framework}. The most sophisticated 
approaches for optimal trading including order book dynamics are based on continuous and 
martingale assumptions \citep{citeulike:6572400} or on Poisson-like point processes \citep{GLFT}.

\begin{figure}[!h]
  \centering
  \includegraphics[width=0.9\textwidth]{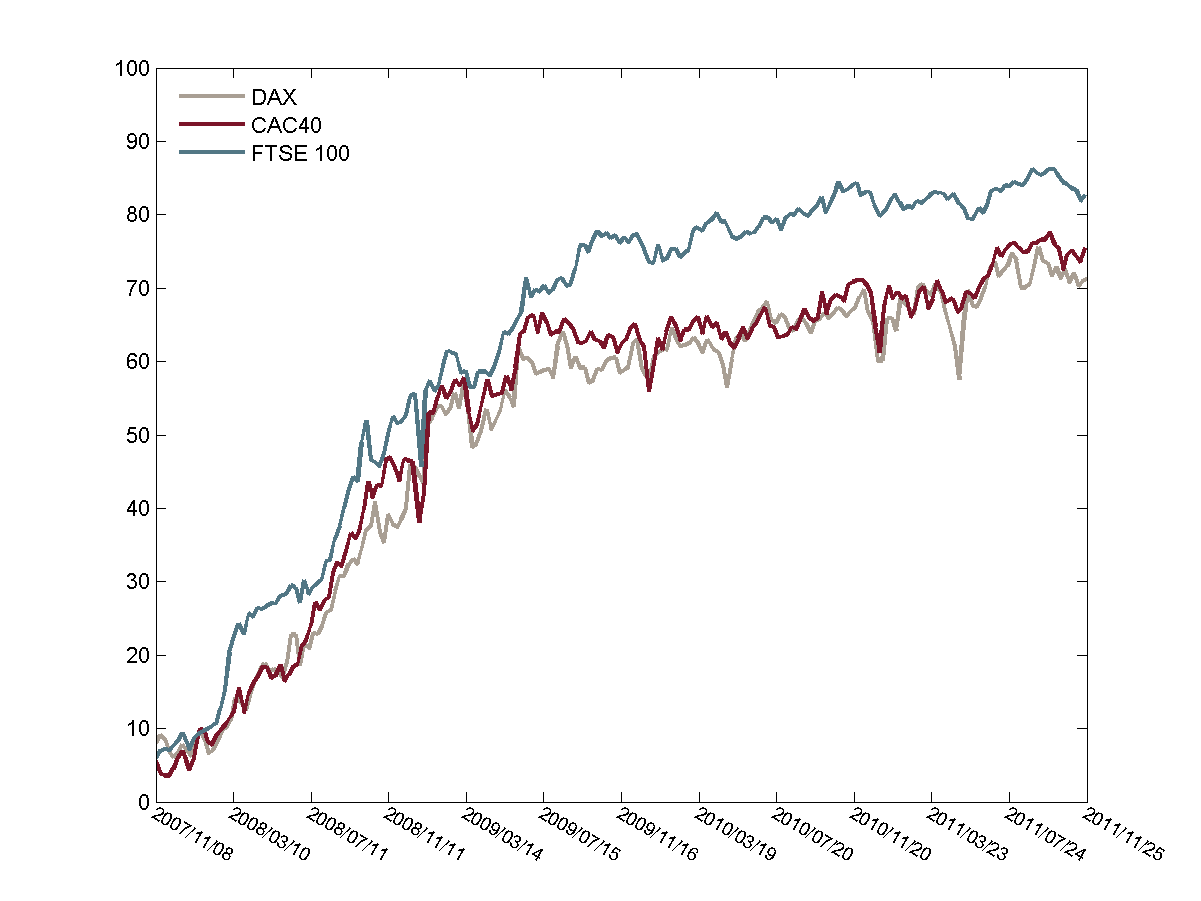}
  \caption{Fragmentation of European markets: the market share of the primary market decreases continuously \rred{after}
 the entry in force of the MiFI Directive; the use of an SOR is mandatory \rred{for accessing} 
the liquidity of the whole market. This graph monitors the \emph{normalized entropy} of 
the fragmentation: if the market shares (summing to $1$) over $K$ exchanges are $q_1,\ldots,q_K$, 
the indicator is $- \sum_k q_k\ln q_k$ renormalized so that its maximum is 100\% (i.e. divided by $\ln(K)$).}
  \label{fig:frag}
\end{figure}

On another hand, focusing on the optimality of very-short-term trading strategies (such as
 \emph{Smart Order Routing}) let us build optimal tactics with assumptions in accordance with 
\rred{recetn high-level} views on order book dynamics.
Smart Order Routers (SOR) are software devices dedicated to splitting an order across all 
available trading venues \rred{in order} to obtain the desired quantity as fast as possible 
implementing a so-called \emph{liquidity capturing scheme}.
With the rise of \rred{fragmented electronic equity markets} (see Figure \ref{fig:frag}), 
it is impossible to access  more than 60\% of European liquidity without a SOR.

Optimal policies for SOR have been proposed in \cite{citeulike:5177512} and \cite{citeulike:7500879}.
The latter used censored statistics to estimate liquidity available and to build an optimization 
framework \rred{on} top of it; the former built a stochastic algorithm and proved that it 
asymptotically converges to a state \rred{that minimizes} a given criterion.

To \rred{get a feel for} the methodology associated with the stochastic algorithm viewpoint, 
just consider the following optimization problem.

\paragraph{Optimal liquidity seeking: the expected fast end \rred{criterion}.}
To define the \rred{criterion to be optimized}, first assume that $K$ visible order books are available 
(for instance BATS, Chi-X, Euronext, Turquoise for an European stock). At time $t$, 
a buy order of size $V_t$ has to be split over the order books according to a key 
$(r_1,\ldots,r_K)$, given that on the $k$th order book:
\begin{itemize}
\item the resting quantity `cheaper' than a given price $S$ is $I^k_t$ 
(i.e. the quantity at the bid side posted at a price higher or equals to $S$, or at a lower price at the ask);
\item the incoming flow of sell orders consuming the resting quantity at prices cheaper 
than $S$ follows a Poisson process $N_t^k$ of intensity $\lambda^k$, i.e.
$$\mathbb{E}( N_{t+\delta t}^k - N^k_t) = \delta t\cdot \lambda^k;$$
\item the waiting time on the $k$th trading destination to consume a volume $v$ added on the $k$th
 trading destination at price $S$ in $t$ is \rred{denoted} $\Delta T_t^k(v)$; it is implicitly defined by:
$$\Delta T_t^k(v)=\arg\min_\tau \{ N_{t+\tau}^k - N^k_t \geq I_t^k + v\}.$$
\end{itemize} 

Assuming that there is no specific toxicity in available trading platforms, a trader would like to split an 
incoming order at time $t$ of size $V_t$ according to an allocation key $(r^1,\ldots,r^K)$ 
\rred{in order to minimize the waiting time criterion. Thus}:
\begin{equation}
  \label{eq:10}
  {\cal C}(r^1,\ldots,r^K) = \mathbb{E} \max_k\{ \Delta T_t^k(r^k V_t)\}.
\end{equation}

\rred{This} means that the trader \rred{aims at optimizing} the following process:
\begin{enumerate}
\item[(1)] an order of size $V_{\tau(u)}$ is to be split at time $\tau(u)$, 
the set of order arrival times \rred{being} ${\cal S}=\{\tau(1),\ldots ,\tau(n),\ldots\}$;
\item[(2)] it is split over the $K$ available trading venues thanks to an `allocation key', 
$\mathbf{R}=(r^1,\ldots,r^K)$: \rred{a portion} $r^kV_{\tau(u)}$ is sent to the $k$th order 
book (all the quantity is spread, i.e. $\sum_{k\leq K} r_k=1$);
\item[(3)] the trader waits the time needed to consume all the sent orders.
\end{enumerate}

The criterion ${\cal C}(\mathbf{R})$ defined in (\ref{eq:10}) reflects the fact 
that the faster the allocation key \rred{lets us} obtain liquidity, the better: 
the obtained key is well suited for a \emph{liquidity-seeking algorithm}.

First \rred{denote by} $k^*_t(\mathbf{R})$ the last trading destination to consume the order sent at $t$:
$$k^*_t(\mathbf{R}) = \arg\max_k \{ \Delta T_t^k(r^k V_t)\}.$$
\rred{A} gradient approach to \rred{minimizing} ${\cal C}(\mathbf{R})$ \rred{means we must}
compute $\partial \Delta T_t^u(r^u V_t)/\partial r^k$ for any pair $(k,u)$. To 
respect the constraint, just replace an arbitrary $r^\ell$ by $1-\sum_{u\neq \ell} r^u$. \rred{Consequently},
$$\df{\Delta T_t^u(r^u V_t)}{r^k}=\df{\Delta T_t^k(r^k)}{r^k} \,\one_{k^*_t(\mathbf{R})=k} + %
                                  \df{\Delta T_t^\ell(r^\ell)}{r^k} \,\one_{k^*_t(\mathbf{R})=\ell}  $$
where $\one$ is a delta function.
\rred{With} the notation $\Delta N^k_t = N^k_t - N^k_{t-}$, we can write that any allocation key $\mathbf{R}$ such that, 
for any pair $(\ell, k)$,
\begin{equation}
  \label{eq:11}
  \mathbb{E}\left( V_t\cdot D^k_t(r^k)\cdot \one_{k^*_t(\mathbf{R})=k}\right) = %
  \mathbb{E}\left( V_t\cdot D^\ell_t(r^\ell)\cdot \one_{k^*_t(\mathbf{R})=\ell}\right)
\end{equation}
where
$$D_t^k(r^k)=\frac{1}{\Delta N^k_{t+\Delta T_t^k(r^k)}}\,\one_{\left(\Delta N^k_{t+\Delta T_t^k(r^k)}>0\right)}$$
is a potential minimum for the criterion ${\cal C}(\mathbf{R})$ (the \rred{proof of this result} will not be provided here).
Equation (\ref{eq:11}) \rred{can also be written as}:
$$\mathbb{E}\left( V_t\cdot D^k_t(r^k)\cdot \one_{k^*_t(\mathbf{R})=k}\right) = \frac{1}{K}\sum_{\ell=1}^K %
  \mathbb{E}\left( V_t\cdot D^\ell_t(r^\ell)\cdot \one_{k^*_t(\mathbf{R})=\ell}\right) .$$

It can be shown (see \cite{citeulike:6053468} for generic results of this kind) that the 
asymptotic solutions of the following stochastic algorithm on the allocation weights through time 
\begin{eqnarray}
  \nonumber
  \forall k,\,r^k(n+1) &=& r^k(n) - \gamma_{k+1} \,\left( V_{\tau(n)}\cdot D^k_{\tau(n)}(r^k(n))\cdot \one_{k^*_{\tau(n)}(\mathbf{R}(n))=k} - \phantom{\sum_{\ell=1}^K}\right.\\
  \label{eq:12} &&\left.\qquad\qquad\qquad 
 \frac{1}{K}\sum_{\ell=1}^K V_{\tau(n)}\cdot D^\ell_{\tau(n)}(r^\ell(n))\cdot 
\one_{k^*_{\tau(n)}(\mathbf{R}(n))=\ell} \right)\qquad~
\end{eqnarray}
\rred{minimize} the expected fast end criterion ${\cal C}(\mathbf{R})$, \rred{provided there are strong enough 
ergodicity assumptions on the $(V,(N^k)_{1\leq k\leq K},(I^k)_{1\leq k\leq K})$-multidimensional process}.

Qualitatively, \rred{we read  this update rule to mean that} if a trading venue $k$ demands more time 
to execute the fraction of the volume that it receives (taking into account the \rred{combination} of $I$ and $N$) 
than the average waiting time on all venues, \rred{then the fraction $r^k$ of the orders to send to $k$ 
has to be decreased for future use}.

\section{Perspectives and Future Work}\label{lehalle_sec5}

The needs of intra-day trading practitioners are currently focused on optimal execution and trading risk control.
\rred{Certainly} some improvements \rred{on} what is actually available \rred{have been} proposed by academics, 
\rred{in particular}:
\begin{itemize}
\item provide optimal trading trajectories taking into account \emph{multiple trading destinations} 
and different type of orders: liquidity-providing (i.e. limit) ones and liquidity-consuming (i.e. market) ones;
\item the \emph{analysis of trading performances} is also an important topic; models 
are needed to understand what part of the performance and risk are due to the planned scheduling, 
the interactions with order books, the market impact and the market moves;
\item \emph{stress testing}: before \rred{executing} a trading algorithm \rred{in} real markets, \rred{we must} 
understand its \rred{dependence on} different market conditions, from volatility or momentum to bid--ask spread or 
trading frequency. The study of the `Greeks' of the payoff of a trading algorithm is not straightforward 
since it is inside a closed loop of liquidity: its `psi' should be its derivative with respect to the bid--ask spread, 
its `phi' with respect to the trading frequency, and its `lambda' with respect to the liquidity available 
in the order book.

For the special case of portfolio \rred{liquidity} studied in this chapter 
(using the payoff $ {\tilde C}_\lambda$ defined by equality~(\ref{eq:8})), 
these trading Greeks would be:
$$\Psi=\left(\frac{\partial {\tilde C}_\lambda}{\partial \psi_\ell}\right)_{1\leq \ell\leq N},\quad %
\Phi= \frac{\partial {\tilde C}_\lambda}{\partial N},\quad %
\Lambda = \frac{\partial {\tilde C}_\lambda}{\partial \kappa} .$$
\end{itemize}

\rred{Progress} in \rred{the above} three directions will provide a better understanding of the price formation 
process and the whole cycle of asset allocation and hedging, taking into account execution costs, 
closed loops with the markets, and portfolio trajectories at any scales.

\paragraph{Acknowledgments}
Most of the data and graphics used here come from the work of Cr\'edit Agricole Cheuvreux 
Quantitative Research group.

%

%
%

\begin{thebibliography}{99}

\bibitem[A\"{i}t-Sahalia and Jacod, 2007]{AITJAC07}
A\"{i}t-Sahalia, Y. and Jacod, J. (2007).
\newblock {Volatility estimators for discretely sampled L\'{e}vy processes}.
\newblock {\em Annals of Statistics} {\bf 35} 355--392.

\bibitem[Alfonsi et~al., 2010]{citeulike:6615020}
Alfonsi, A., Fruth, A., and Schied, A. (2010).
\newblock {Optimal execution strategies in limit order books with general shape
  functions}.
\newblock {\em Quantitative Finance} {\bf 10} (2) 143--157.

\bibitem[Alfonsi and Schied, 2010]{citeulike:6572400}
Alfonsi, A. and Schied, A. (2010).
\newblock {Optimal execution and absence of price manipulations in limit order
  book models}.
\newblock {\em SIAM J. Finan. Math.} {\bf 1} 490--522.

\bibitem[Almgren et~al., 2005]{citeulike:4325901}
Almgren, R., Thum, C., Hauptmann, E., and Li, H. (2005).
\newblock {Direct estimation of equity market impact}.
\newblock {\em Risk} {\bf 18} 57--62.

\bibitem[Almgren and Chriss, 2000]{OPTEXECAC00}
Almgren, R.~F. and Chriss, N. (2000).
\newblock {Optimal execution of portfolio transactions}.
\newblock {\em Journal of Risk} {\bf 3} (2) 5--39.

\bibitem[Altunata et~al., 2010]{citeulike:6716078}
Altunata, S., Rakhlin, D., and Waelbroeck, H. (2010).
\newblock {Adverse selection vs. opportunistic savings in dark aggregators}.
\newblock {\em Journal of Trading} {\bf 5} 16--28.

\bibitem[Avellaneda and Stoikov, 2008]{avst08}
Avellaneda, M. and Stoikov, S. (2008).
\newblock {High-frequency trading in a limit order book}.
\newblock {\em Quantitative Finance} {\bf 8} (3) 217--224.

\bibitem[Bacry et~al., 2009]{citeulike:7344893}
Bacry, E., Delattre, S., Hoffman, M., and Muzy, J.~F. (2009).
\newblock {Deux mod\`{e}les de bruit de microstructure et leur inf\'{e}rence
  statistique}.
\newblock \rred{Preprint, CMAP, \'Ecole Polytechnique, France}.

\bibitem[Biais et~al., 2005]{RePEc:ide:wpaper:825}
Biais, B., Glosten, L., and Spatt, C. (2005).
\newblock {Market microstructure: a survey of microfoundations, empirical
  results, and policy implications}.
\newblock {\em Journal of Financial Markets} {\bf 2} (8) 217--264.

\bibitem[Bouchard et~al., 2011]{citeulike:5797837}
Bouchard, B., Dang, N.-M., and Lehalle, C.-A. (2011).
\newblock {Optimal control of trading algorithms: a general impulse control
  approach}.
\newblock {\em SIAM J. Financial Mathematics} {\bf 2} 404--438.

\bibitem[Bouchaud et~al., 2002]{citeulike:1618840}
Bouchaud, J.~P., Mezard, M., and Potters, M. (2002).
\newblock {Statistical properties of stock order books: empirical results and
  models}.
\newblock {\em Quantitative Finance} {\bf 2} (4) \rred{251--256}. 

\bibitem[Chakraborti et~al., 2011]{citeulike:5823204}
Chakraborti, A., Toke, I.~M., Patriarca, M., and Abergel, F. (2011).
\newblock {Econophysics review: II. agent-based models}.
\newblock {\em Quantitative Finance} {\bf 11} (7) \rred{ 1013--1041}. 

\bibitem[Cohen et~al., 1981]{citeulike:7604491}
Cohen, K.~J., Maier, S.~F., Schwartz, R.~A., and Whitcomb, D.~K. (1981).
\newblock {Transaction costs, order placement strategy, and existence of the
  bid--ask spread}.
\newblock {\em The Journal of Political Economy} {\bf 89} (2) 287--305.

\bibitem[Cont and De~Larrard, 2011]{citeulike:8531765}
Cont, R. and De~Larrard, A. (2011).
\newblock {Price dynamics in a Markovian limit order book market}.
\newblock {\em Social Science Research Network Working Paper Series}.

\bibitem[Cont et~al., 2010]{citeulike:8318790}
Cont, R., Kukanov, A., and Stoikov, S. (2010).
\newblock {The price impact of order book events}.
\newblock {\em Social Science Research Network Working Paper Series}.

\bibitem[Engle et~al., 2012]{citeulike:4368376}
Engle, R.~F., Ferstenberg, R., and Russell, J.~R. (2012).
\newblock {Measuring and modeling execution cost and risk}.
\newblock {\em The Journal of Portfolio Management} {\bf 38} (2) 14--28.

\bibitem[Foucault and Menkveld, 2008]{FOU06}
Foucault, T. and Menkveld, A.~J. (2008).
\newblock {Competition for order flow and smart order routing systems}.
\newblock {\em The Journal of Finance} {\bf 63} (1) 119--158.

\bibitem[Gabaix et~al., 2006]{citeulike:7360166}
Gabaix, X., Gopikrishnan, P., Plerou, V., and Stanley, H.~E. (2006).
\newblock {Institutional investors and stock market volatility}.
\newblock {\em Quarterly Journal of Economics} {\bf 121} (2) m461--504.

\bibitem[Ganchev et~al., 2010]{citeulike:7500879}
Ganchev, K., Nevmyvaka, Y., Kearns, M., and Vaughan, J.~W. (2010).
\newblock {Censored exploration and the dark pool problem}.
\newblock {\em Commun. ACM} {\bf 53} (5) 99--107.

\bibitem[Gatheral, 2010]{citeulike:5177397}
Gatheral, J. (2010).
\newblock {No-dynamic-arbitrage and market impact}.
\newblock {\em Quantitative Finance} {\bf 10} (7) \rred{749--759}.  

\bibitem[Gatheral and Schied, 2012]{citeulike:10363463}
Gatheral, J. and Schied, A. (2012).
\newblock {Dynamical models of market impact and algorithms for order
  execution}.
\newblock {\em Handbook of Systemic
  Risk}. Fouque, J.-P. and Langsam, J., (eds). Cambridge University Press.

\bibitem[Gu\'{e}ant et~al., 2011]{GLFT}
Gu\'{e}ant, O., Lehalle, C.~A., and Fernandez-Tapia, J. (2011).
\newblock {Optimal execution with limit orders}.
\newblock {\em Working paper}. 

\bibitem[Hayashi and Yoshida, 2005]{YOSHI05}
Hayashi, T. and Yoshida, N. (2005).
\newblock {On covariance estimation of non-synchronously observed diffusion
  processes}.
\newblock {\em Bernoulli} {\bf 11} (2) 359--379.

\bibitem[Hewlett, 2006]{citeulike:7012187}
Hewlett, P. (2006).
\newblock {Clustering of order arrivals, price impact and trade path
  optimisation}.
\newblock In {\em Workshop on Financial Modeling with Jump processes}. Ecole
  Polytechnique.

\bibitem[Ho and Stoll, 1981]{RePEc:eee:jfinec:v:9:y:1981:i:1:p:47-73}
Ho, T. and Stoll, H.~R. (1981).
\newblock {Optimal dealer pricing under transactions and return uncertainty}.
\newblock {\em Journal of Financial Economics} {\bf 9} (1) 47--73.

\bibitem[Jacod, 1996]{VQ96}
Jacod, J. (1996).
\newblock La
  variation quadratique moyenne du brownien en pr\'{e}sence d'erreurs
  d'arrondi.
\newblock In  {\em {Hommage \`a P. A. Meyer et J. Neveu}},  
\newblock Asterisque, {\bf 236}.

\bibitem[Kirilenko et~al., 2010]{citeulike:8676220}
Kirilenko, A.~A., Kyle, A.~P., Samadi, M., and Tuzun, T. (2010).
\newblock {The flash crash: the impact of high frequency trading on an
  electronic market}.
\newblock {\em Social Science Research Network Working Paper Series}.

\bibitem[Kyle, 1985]{citeulike:3320208}
Kyle, A.~P. (1985).
\newblock {Continuous auctions and insider trading}.
\newblock {\em Econometrica} {\bf 53} (6) 1315--1335.

\bibitem[Large, 2007]{citeulike:7012175}
Large, J. (2007).
\newblock {Measuring the resiliency of an electronic limit order book}.
\newblock {\em Journal of Financial Markets} {\bf 10} (1) 1--25.

\bibitem[Lasry and Lions, 2007]{citeulike:3614137}
Lasry, J.-M. and Lions, P.-L. (2007).
\newblock {Mean field games}.
\newblock {\em Japanese Journal of Mathematics} {\bf 2} (1) 229--260.

\bibitem[Lehalle, 2009]{citeulike:5094012}
Lehalle, C.-A. (2009).
\newblock {Rigorous strategic trading: balanced portfolio and mean-reversion}.
\newblock {\em The Journal of Trading} {\bf 4} (3) 40--46.

\bibitem[Lehalle et~al., 2010]{citeulike:7621540}
Lehalle, C.-A., Gu\'{e}ant, O., and Razafinimanana, J. (2010).
\newblock {High frequency simulations of an order book: a two-scales approach}.
\newblock In  {\em Econophysics of Order-Driven Markets}, Abergel, F., Chakrabarti, 
B.~K., Chakraborti, A., and Mitra, M.,
(eds), New Economic Windows.
  Springer.

\bibitem[Lelong, 2011]{citeulike:6053468}
Lelong, J. (2011).
\newblock {Asymptotic normality of randomly truncated stochastic algorithms}.
\newblock {\em ESAIM: Probability and Statistics} (forthcoming).  

\bibitem[Lillo et~al., 2003]{NAT03}
Lillo, F., Farmer, J.~D., and Mantegna, R. (2003).
\newblock {\rred{Master curve for price--impact function}}.
\newblock {\em Nature}, {\bf 421}, \rred{129--130}.  

\bibitem[Markowitz, 1952]{citeulike:571949}
Markowitz, H. (1952).
\newblock {Portfolio selection}.
\newblock {\em The Journal of Finance} {\bf 7} (1) 77--91.

\bibitem[Menkveld, 2010]{citeulike:8423311}
Menkveld, A.~J. (2010).
\newblock {High frequency trading and the new-market makers}.
\newblock {\em Social Science Research Network Working Paper Series}.

\bibitem[Muniesa, 2003]{MUN03}
Muniesa, F. (2003).
\newblock {\em {Des march\'{e}s comme algorithmes: sociologie de la cotation
  \'{e}lectronique \`{a} la Bourse de Paris}}.
\newblock PhD thesis, Ecole Nationale Sup\'{e}rieure des Mines de Paris.

\bibitem[Pag\`{e}s et~al., 2012]{citeulike:5177512}
Pag\`{e}s, G., Laruelle, S., and Lehalle, C.-A. (2012).
\newblock {Optimal split of orders across liquidity pools: a stochatic
  algorithm approach}.
\newblock {\em SIAM Journal on Financial Mathematics} \rred{{\bf 2} (1) 1042--1076.} 

\bibitem[Predoiu et~al., 2011]{citeulike:8531791}
Predoiu, S., Shaikhet, G., and Shreve, S. (2011).
\newblock {Optimal execution of a general one-sided limit-order book}.
\newblock {\em SIAM Journal on Financial Mathematics} {\bf 2} 183--212.

\bibitem[Robert and Rosenbaum, 2011]{citeulike:8317402}
Robert, C.~Y. and Rosenbaum, M. (2011).
\newblock {A new approach for the dynamics of ultra-high-frequency data: the
  model with uncertainty zones}.
\newblock {\em Journal of Financial Econometrics} {\bf 9} (2) 344--366.

\bibitem[Shiryaev, 1999]{citeulike:1681881}
Shiryaev, A.~N. (1999).
\newblock {\em {Essentials of Stochastic Finance: Facts, Models, Theory}}, 1st edition.
\newblock World Scientific Publishing Company.

\bibitem[Smith et~al., 2003]{farmer03a}
Smith, E., Farmer, D.~J., Gillemot, L., and Krishnamurthy, S. (2003).
\newblock {Statistical theory of the continuous double auction}.
\newblock {\em Quantitative Finance} {\bf 3} (6) 481--514.

\bibitem[Wyart et~al., 2008]{Bouchaud06}
Wyart, M., Bouchaud, J.-P., Kockelkoren, J., Potters, M., and Vettorazzo, M.
  (2008).
\newblock {Relation between bid--ask spread, impact and volatility in double
  auction markets}.
\newblock \rred{{\em Quantitative finance} {\bf 8} 41--57.}

\bibitem[Zhang et~al., 2005]{scales05}
Zhang, L., Mykland, P.~A., and Sahalia, Y.~A. (2005).
\newblock {A tale of two time scales: determining integrated volatility with
  noisy high-frequency data}.
\newblock {\em Journal of the American Statistical Association} {\bf 100} (472) \rred{1394--1411}. 

\end{thebibliography}
  \end{document}